\documentclass[11pt]{article}
\usepackage[utf8]{inputenc}

\usepackage{fullpage}

\usepackage{hyperref}
\usepackage{amsmath}
\usepackage{amsthm}
\usepackage{cleveref}
\usepackage{amssymb}
\usepackage{mathtools}
\usepackage[linesnumbered]{algorithm2e}
\usepackage{thmtools}
\usepackage{thm-restate}

\newtheorem{theorem}{Theorem}

\newtheorem{lemma}[theorem]{Lemma}
\newtheorem{claim}[theorem]{Claim}
\newtheorem{definition}[theorem]{Definition}
\newtheorem{proposition}[theorem]{Proposition}
\newtheorem{observation}[theorem]{Observation}
\newtheorem{remark}[theorem]{Remark}

\renewenvironment{proof}{\noindent\bf{Proof.}\rm}{\hfill$\blacksquare$\bigskip}

\newcommand{\items}{\mathcal{M}}

\usepackage[]{color-edits}
\addauthor{UF}{red}

\addauthor{GB}{blue}

\title{On fair allocation of indivisible goods to submodular agents}

\author{Gilad Ben Uziahu\thanks{Weizmann Institute ---  E-mail: \texttt{gilad.ben-uziahu@weizmann.ac.il}.}\;\, and Uriel Feige\thanks{Weizmann Institute ---  E-mail: \texttt{uriel.feige@weizmann.ac.il}.}}

\begin{document}

\maketitle

\begin{abstract}
We consider the problem of fair allocation of indivisible goods to agents with submodular valuation functions, where agents may have either equal entitlements or arbitrary (possibly unequal) entitlements. We focus on share-based fairness notions, specifically, the maximin share (MMS) for equal entitlements and the anyprice share (APS) for arbitrary entitlements, and design allocation algorithms that give each agent a bundle of value at least some constant fraction of her share value. For the equal entitlement case (and submodular valuations), Ghodsi, Hajiaghayi, Seddighin, Seddighin, and Yami [EC 2018] designed a polynomial-time algorithm for $\frac{1}{3}$-maximin-fair allocation. We improve this result in two different ways. We consider the general case of arbitrary entitlements, and present a polynomial time algorithm that guarantees submodular agents $\frac{1}{3}$ of their APS. For the equal entitlement case, we improve the approximation ratio and obtain $\frac{10}{27}$-maximin-fair allocations. Our algorithms are based on designing strategies for a certain bidding game that was previously introduced by Babaioff, Ezra and Feige [EC 2021].

\end{abstract}

\section{Introduction}

We study the problem of fair allocation of a set $\mathcal{M}$ of $m$ indivisible items to a set $\mathcal{N}$ of $n$ agents, where each agent $i$ has a monotone non-negative submodular valuation function $v_{i}:2^{\mathcal{M}}\to\mathbb{R}_{\geq0}$ (the definition of submodularity appears in \Cref{sec: Classes of valuation functions}). In fair allocation settings, agents do not pay for the items. Instead, agents have arbitrary, possibly unequal entitlements to the items. Specifically, each agent $i$ has an individual entitlement $0<b_i\leq1$, and the entitlements sum up to~1 ($\sum_{i=1}^n b_i = 1$). We focus on share-based fairness notions, specifically, the maximin share (MMS) for equal entitlements~\cite{Budish11} and the anyprice share (APS) for arbitrary entitlements~\cite{BEF21} (definitions of these notions appear in Section~\ref{sec:basic definitions}). We design allocation algorithms that give each agent a bundle of value at least some constant fraction of her share value.

{The problem of fairly allocating indivisible items has been extensively studied, with various settings of the problem considered (items might be goods or chores, agents may have equal or unequal entitlements), and different fairness criteria explored, such as envy-based notions and share-based principles. See for example \cite{Aziz_Survey22, Amanatidis_survey22} and references therein.}

The problem of allocating indivisible items to agents arises naturally in the real-world. We present one such example that illustrates aspects addressed in our work (unequal entitlement, no payments, non-additive valuation functions). The NBA draft is an annual event of the National Basketball Association (NBA) in which new eligible basketball players (typically graduate college players) are allocated to NBA teams. 
The allocation mechanism is a picking sequence composed of two rounds. In each round, each team in its turn picks a player among the eligible players. In this example, teams correspond to the agents, and basketball players correspond to the items.
The teams do not have equal entitlement to the players. Teams of poorer performance in the previous season have higher entitlement that those of better performance (a policy that tries to maintain the competitiveness of the teams). This unequality of entitlement is reflected in the allocation mechanism, by having teams with higher entitlement pick earlier than teams of lower entitlement, in each of the rounds of the picking sequence. (In practice, the allocation mechanism is somewhat more complicated than described above, but these additional complications are not relevant to our presentation, and hence omitted.) 
Teams do not pay for the right to pick a player. (They will of course later pay the salary of the player, but these monetary aspects are only an aspect that determines how desirable the player is for the team, and are not part of the allocation processes.)
The interests of teams do not seem to fit a model of an additive valuation function. For example, 
it is likely that the combined value for a team of two players that play in the ``center" position is smaller than the sum of values of the individual players.

{
\subsection{Fairness notions}

In our paper, we focus on notions of \emph{fairness} known as \emph{share-based} notions. In share-based fairness, each agent cares only about her own bundle in the allocation, and expects its value to reach at least a certain target value.
One such fairness notion is the \emph{maximin share}, which was introduced by Budish \cite{Budish11}. The \emph{maximin share} (abbreviated as \emph{MMS}) of an agent is defined to be the maximum value she can ensure for herself if she were to partition the goods into $n$ bundles and then receive a minimum valued bundle. A \emph{maximin fair} allocation is an allocation in which each agent gets a bundle that she values at least as her maximin share.
}

The maximin share notion is applicable when agents have equal entitlement. A notion of fairness for the case of arbitrary entitlements was presented by Babaioff, Ezra, and Feige \cite{BEF21}, and is referred to as the \emph{AnyPrice share} (abbreviated as APS). See Definition~\ref{def:APS}.
In the special case of equal entitlements, the AnyPrice share of an agent is at least as large as her \emph{Maximin share}, and sometimes strictly larger.


\subsection{Notions of approximation}

In the context of the notions of the MMS and APS, there are two different tasks that involve approximations.

\begin{itemize}

\item Approximating the value of the MMS (or APS) of an agent. Both the MMS and APS of an agent are NP-hard to compute even if the valuation function is additive, in which case computing the exact value of the MMS is strongly NP-hard~\cite{Woeginger97}, whereas computing the exact value of the APS is weakly NP-hard~\cite{BEF21}. For submodular valuation functions (a class that is considered in this paper), computing the MMS and the APS is APX-hard. See Section~\ref{sec:APX} for more details.

\item Approximating a fair allocation (maximin-fair allocation, AnyPrice-fair allocation, etc.). For $\alpha\in(0,1)$, we say that an allocation is $\alpha$-\emph{maximin-fair }(resp. $\alpha$-\emph{AnyPrice-fair)} if it gives every agent at least an $\alpha$ fraction of her MMS (resp. APS). Kurokawa, Procaccia, and Wang \cite{KPW18} showed that for every $n\geq3$, there exists an instance with $n$ additive agents, such that no \emph{maximin-fair} allocation exist, i.e., for each allocation, there exist an agent which gets a bundle she values strictly less than her MMS. As the APS of an agent is at least as large as her MMS, there are instances with additive valuations and no APS-allocation.
\end{itemize}

In our paper we will focus on the latter task, approximating MMS-fair (APS-fair) allocations.

\subsection{Classes of valuation functions}
\label{sec: Classes of valuation functions}

Throughout this paper we assume that valuation functions are {\em normalized} (the value of the empty set is~0) and {\em monotone} ($v(S) \le v(T)$ for $S \subset T$).

Lehman, Lehman, and Nissan \cite{DBLP:journals/geb/LehmannLN06}introduce a hierarchy of families of valuation functions, and two prominent members of this hierarchy are \emph{Submodular} and \emph{XOS} valuations, as defined below.

\begin{definition}
\textbf{(Submodular valuation) }a valuation function $v\colon2^{\mathcal{M}}\to\mathbb{R}_{\geq0}$ is submodular if the following (equivalent) conditions hold: 
\begin{itemize}
\item $\forall S,T\subseteq\mathcal{M}$ we have $v(S)+v(T)\geq v(S\cup T)+v(S\cap T)$
\item $\forall S,T\subseteq\mathcal{M}$ with $S\subseteq T$, and for any
$j\in\mathcal{M\setminus}T$ we have $v(S\cup\{j\})-v(S)\geq v(T\cup\{j\})-v(T)$
\end{itemize}
\end{definition}
\begin{definition}
\textbf{(XOS valuation)} a valuation function $v\colon2^{\mathcal{M}}\to\mathbb{R}_{\geq0}$
is XOS (also referred to as {\em fractionally subadditive}) if there exist a finite set of \textbf{additive} valuations $\{v_{1},v_{2},\dots,v_{k}\}$ such that
\[
\forall T\in\mathcal{M},\quad v(T)=\max_{j\in[k]}v_{j}(T)
\]

\end{definition}


As shown in \cite{DBLP:journals/geb/LehmannLN06}, the hierarchy of these classes is as follows:
\[
Additive\subsetneq Submodular\subsetneq XOS
\]

{
Let us briefly discuss the representation of valuation functions.
The explicit representation of a valuation function requires exponential space in $m$ (its domain size is $2^{m}$). 
Consequently, as described in \cite{DBLP:journals/geb/LehmannLN06}, one typically assumes query access to valuation functions, rather than having an explicit representation for them.
Our paper will focus on the value queries model (i.e., the function is implicitly given through a value oracle). In this model, a query is a set of items, and the answer is the value of the function on this set of items. We assume that each such query takes unit time. Consequently, polynomial allocation algorithms may make only polynomially many value queries to the underlying valuation functions. 
}

\subsection{Our main results}

Our main results concern the existence (and polynomial time computability) of approximate MMS-fair (APS-fair) allocations in the equal entitlements (arbitrary entitlements) case for submodular agents. Previously, for the equal entitlement case, Ghodsi, Hajiaghayi, Seddighin, Seddighin and Yami \cite{DBLP:conf/sigecom/GhodsiHSSY18} designed a polynomial-time algorithm for $\frac{1}{3}$-\emph{maximin-fair} allocations, and designed instances in which in every allocation, at least one agent gets a bundle of value not larger than a $\frac{3}{4}$ fraction of her MMS. 

Our results are based on a bidding game mechanism presented by \cite{BEF21}, where each agent gets an initial budget, and in every round, the highest bidder gets to choose an item. We show here that in this bidding game, a submodular agent has a bidding strategy that guarantees at least a $\frac{1}{3}$ fraction of her APS in the arbitrary entitlements case. 
In the case of equal entitlements, we show that with a slight modification of the bidding game (named as the \emph{altruistic version} of the bidding game, {see \Cref{sec:Approximated MMS-fair existance}}), there is a bidding strategy for a submodular agent that guarantees at least a $\frac{10}{27}$ fraction of her MMS.

\begin{restatable}{rethm}{APSbidding}\label{thm:1/3_APS_guarantee}
 Consider the bidding game described above, and an agent $p$ with a submodular valuation function and entitlement $b_p$. Setting $\rho={\frac{1}{3-2b_p} > \frac{1}{3}}$, a bidding strategy referred to as $proportional(\rho)$ guarantees agent $p$ a value of at least $\rho \cdot APS_p$. (In the case of equal entitlements, this gives $\rho=\frac{n}{3n-2}$.)
\end{restatable}

\begin{restatable}{rethm}{altruistic}
\label{thm:equal-10/27}
    Consider the altruistic version of the bidding game in the equal entitlement case. Every agent with a submodular valuation that uses a bidding strategy referred to as the proportional bidding strategy is guaranteed to get a bundle of value at least a $\rho = \frac{10}{27} + \Omega(\frac{1}{n}) > 0.37037$ fraction of her MMS. 
\end{restatable}

The bidding strategies used in our two main theorems require computing the APS (or MMS) of the agents, tasks which are APX-hard. Nevertheless, known techniques~\cite{DBLP:conf/sigecom/GhodsiHSSY18} allow us to deduce the following corollary:

\begin{restatable}{recor}{PolySubmodular}
\label{cor:polyTimeSubmodular}
There is a polynomial time $\frac{1}{3}$-APS algorithm for submodular valuations with arbitrary entitlements. For the equal entitlement case, there is a polynomial time $\frac{10}{27}$-MMS algorithm for submodular valuations.
\end{restatable}

Our results concerning submodular valuations can be combined with previous results of \cite{BEF21} that concern bidding strategies for agents with subclasses of submodular valuations, namely additive valuations and unit demand valuations. This results in allocation algorithms for setting with submodular valuations, in which agents that have valuations coming from simple sub-classes of submodular valuations get improved guarantees.



\begin{restatable}{recor}{PolyEnsamble}
\label{cor:submodular, additive, and unit demand algorithm}
There is a polynomial time allocation algorithm, which simultaneously guarantees for submodular agents $\frac{1}{3}$-APS, for additive agents $\frac{3}{5}$-APS, and for Unit-demand agents $1$-APS.
\end{restatable}

A class of valuations that is more general than submodular valuations is XOS valuations. 
A natural question concerning the bidding game is whether agents with XOS valuations have strategies that guarantee a constant fraction of their APS.  The answer for this question is negative.

\begin{restatable}{reprop}{XosHardness}
\label{prop:XOS_hardness}
There is no bidding strategy that guarantees a constant fraction of the MMS to an agent with an XOS valuation function (not even in the case of equal entitlements).
\end{restatable}

\subsection{Basic definitions}
\label{sec:basic definitions}

\begin{definition}
\textbf{(Maximin share (MMS))} Consider an allocation instance with a set $\mathcal{M}=\{e_{1},\dots,e_{m}\}$ of $m$ items and a set $\mathcal{N}=\{1,\dots,n\}$ of $n$ agents, where each agent $i$ has an individual non-negative valuation function $v_{i}\colon2^{\mathcal{M}}\to\mathbb{R}_{\geq0}$. Then the maximin share of agent $i$, denoted by $MMS_{i}$, is the maximum over all $n$-partitions of $\mathcal{M}$, of the minimum value under $v_{i}$ of a bundle in the $n$-partition
\[
MMS_{i}=\max_{A_{1},A_{2},\dots,A_{n}\in P_{n}(\mathcal{M})}\min_{j}v_{i}(A_{j})
\]

(where $P_{n}(\mathcal{M})$ is the set of all partitions of $\mathcal{M}$ to $n$ pairwise disjoint sets)
\end{definition}
\begin{definition}
\label{def:APS}
\textbf{(AnyPrice share)} Consider a setting in which agent $i$ with valuation $v_{i}$ has entitlement $b_{i}$ to a set of indivisible items $\mathcal{M}$. The AnyPrice share (APS) of agent $i$, denoted by $AnyPrice(b_{i},v_{i},\mathcal{M})$, is the value she can guarantee herself whenever the items in $\mathcal{M}$ are adversarially priced with non-negative prices that sum up to 1, and she picks her favorite affordable bundle. More formally, if $P=\left\{ (p_{1},\dots,p_{m})|\sum p_{j}=1,\text{ and }\forall j,p_{j}\geq0\right\}$ is the set of all possible pricing of $\mathcal{M}$, then the definition of the APS is:
\[
AnyPrice(b_{i},v_{i},\mathcal{M})=\min_{(p_{1},p_{2},\dots,p_{m})\in P}\max_{S\subseteq\mathcal{M}}\left\{ v_{i}(S) \mid \sum_{j\in S}p_{j}\leq b_{i}\right\}
\]

When $\mathcal{M}$ and $v_{i}$ are clear from context we denote the APS share of an agent $i$ with entitlement $b_{i}$ by $AnyPrice(b_{i})$, instead of $AnyPrice(b_{i},v_{i},\mathcal{M})$.
\end{definition}
$\quad$

As shown in \cite{BEF21}, the AnyPrice share has the following equivalent definition.
\begin{definition}
\label{def:dual}
\textbf{(AnyPrice share dual definition)} Consider a setting in which agent $i$ with valuation $v_{i}$ has entitlement $b_{i}$ to a set of indivisible items $\mathcal{M}$. The AnyPrice share of $i$, denoted by $AnyPrice(b_{i},v_{i},\mathcal{M})$, is the maximum value $z$ she can get by coming up with nonnegative weights $\left\{ \lambda_{T}\right\} _{T\in\mathcal{M}}$ that total to 1 (a distribution over sets), such that any set $T$ of value below $z$ has a weight of zero, and any item appears in
sets of a total weight at most $b_{i}$:
\[
AnyPrice(b_{i},v_{i},\mathcal{M})=\max z
\]
 subject to the following set of constraints being feasible for $z$:
\begin{itemize}
\item $\sum_{T\subseteq\mathcal{M}}\lambda_{T}=1$
\item $\lambda_{T}\geq0,\forall T\subseteq\mathcal{M}$
\item $\lambda_{T}=0,\forall T\subseteq\mathcal{M}\text{ s.t }v_{i}(T)<z$
\item $\sum_{T:j\in T}\lambda_{T}\leq b_{i},\forall j\in\mathcal{M}$
\end{itemize}
\end{definition}

\subsection{Related work}


There are several different approaches trying to define fairness criteria for allocation of items. One approach concerns elimination (or minimization) of envy among agents. An allocation is {\em envy-free}  if no agent strictly prefers a bundle of another agent over her own bundle {\cite{Foley67}}. Envy-free allocations exist in setting with divisible items, but need not exist in settings with indivisible items (e.g., when there are fewer agents than items). Consequently, various relaxations of the envy-free property have been introduced, among them EF1 {(\cite{LiptonEF1},\cite{Budish11})} and EFX {\cite{MaximumNashWelfare}}. In this work we do not consider envy-based fairness notions.

Perhaps the first share-based fairness notion to have been introduced is the proportional share, $b_i \cdot v_{i}(\mathcal{M})$. This notion and various relaxations of it (Prop1) may be appropriate when valuations functions are additive, but is hard to justify for other classes of valuation functions. In this work we are concerned with submodular valuations. For the case of equal entitlements, we consider the maximin share (MMS) \cite{Budish11}, which is the share notion that is most commonly used for allocation of indivisible items to agents with equal entitlements. For the case of arbitrary entitlements, we use the anyprice share (APS) \cite{BEF21}. We remark that there are other notions of shares that have been proposed for settings with unequal entitlements and are not considered in our work. These include the \emph{weighted maximin share} (WMMS) of~\cite{FGT19}, and the \emph{l-out-of-d} share~\cite{DBLP:journals/mor/BabaioffNT21}.

We present here some known approximation results for \emph{MMS-fair} and \emph{APS-fair} allocations:

\begin{itemize}

\item Additive valuations with equal entitlements (approximate MMS-allocations).

\begin{itemize}

\item Impossibility results. As mentioned above, Kurokawa, Procaccia, and Wang \cite{KPW18} were the first to show that for every $n\geq3$, there exists an instance with $n$ additive agents, such that no \emph{maximin-fair }allocation exists. Later, Feige, Sapir, and Tauber \cite{DBLP:journals/corr/abs-2104-04977} showed an example of an instance with $n=3$ agents with additive valuations, where for any allocation, at least one of the agents gets a bundle she values at most $\frac{39}{40}$ of her MMS). 

\item Existence results. \cite{KPW18} showed existence of $\approx\frac{2}{3}$-\emph{maximin-fair }allocation (up to $O(\frac{1}{n})$). Ghodsi et al \cite{DBLP:conf/sigecom/GhodsiHSSY18} showed existence of $\frac{3}{4}$-\emph{maximin-fair} allocations. 

\item Algorithmic results. Garg and Taki \cite{GargTaki21}, presented a polynomial time algorithm that finds a $\frac{3}{4}$-\emph{maximin-fair} allocation, assuming value queries. 

\end{itemize}

\item 
Submodular valuations with equal entitlements. \cite{BarmanKumar17} showed existence and polynomial time computability of $\approx0.21$-MMS fair allocations. Their algorithm is based on a simple round-robin algorithm. Ghodsi et al \cite{DBLP:conf/sigecom/GhodsiHSSY18} showed a polynomial-time algorithm for $\frac{1}{3}$-\emph{maximin-fair} allocations, and examples in which $\rho$-\emph{maximin-fair} allocations do not exist, for any $\rho > \frac{3}{4}$ and any $n \ge 2$ (number of agents).

\item For XOS valuations with equal entitlements, Ghodsi et al \cite{DBLP:conf/sigecom/GhodsiHSSY18} show the existence of $\frac{1}{5}$-MMS allocations, and presented examples in which $\rho$-\emph{maximin-fair} allocations do not exist, for any $\rho > \frac{1}{2}$ and any $n \ge 2$. 


\item Additive valuations with arbitrary entitlements (approximate \emph{AnyPrice-fair} allocations).
\begin{itemize}
\item Impossibility result. The negative results stated for MMS allocations (such as upper bound of $\frac{39}{40}$ for $n = 3$ agents) extend to APS allocations, as the APS is at least as large as the MMS. 
\item There exists a polynomial-time algorithm for computing a $\frac{3}{5}$-APS
allocation, i.e., an algorithm which returns an allocation where each agent gets a bundle she values at least $\frac{3}{5}$ of her AnyPrice share \cite{BEF21}.
\end{itemize}
\end{itemize}

We are not aware of previous work concerning approximate APS-allocations for valuation functions that are submodular.


\section{Proofs of our results}


\subsection{Preliminaries}
\label{sec:preliminaries}

\begin{claim}\label{cl:item_reduce}
Let $I$ be an allocation instance with a set $\items$ of items, let $i$ be an agent with entitlement $0 < b_i < 1$, let $j$ be an agent with entitlement $b_j \le b_i$, and let $e \in \items$ be an arbitrary item. Consider an allocation {instance} $I'$ that differs from $I$ only in that agent $i$ is removed, the entitlements of all other agents are scaled by $\frac{1}{1 - b_i}$ (so that they sum up to~1), and item $e$ is removed. Then the APS of $j$ in $I'$ is at least as large as the APS of $j$ in $I$. Moreover, if $I$ was an instance with equal entitlement, then the same applies to the MMS of $j$
\end{claim}

\begin{proof}
The claim above states that for each agent $j$ with $b_{j}\leq b_{i}$,
if we define $b_{j}^{*}:=\frac{1}{1-b_{i}}\cdot b_{j}$, then:
\[
APS_{j}(\mathcal{M}\setminus\{e\},b_{j}^{*})\geq APS_{j}(\mathcal{M},b_{j})
\]

Let $\{\lambda_{S}S\}$ be a fractional APS  partition for $j$ in the original instance $I$. Let $w=\sum_{S|e\in S}\lambda_{S}$ denote the total weight of bundles in which $e$ participates in the fractional partition. Note that by definition of the APS, $w \le b_j$, and since $b_j \le b_i$, we also have $w \le b_i$. Define the following new weights:
\[
\lambda_{S}^{*}=\begin{cases}
\frac{1}{(1-w)}\cdot\lambda_{S} & \text{ if }e\not\in S\\
0 & \text{ otherwise}
\end{cases}
\]

We show that the new weights induce a fractional APS partition for the new instance $I'$ (with agent $i$ and item $e$ removed). Indeed, every item is in bundles of total weight at most $b^*_j$:
\begin{align*}
\forall e' & \in\mathcal{M}\setminus\{e\},\quad\sum_{S|e'\in S}\lambda_{S}^*\leq\frac{1}{(1-w)}\sum_{S|e'\in S}\lambda_{S}\leq\frac{1}{(1-w)}b_{j}\leq\frac{1}{(1-b_{i})}b_{j}=b_{j}^{*}
\end{align*}

The sum of weights of all bundles is 1:
\[
\sum_{S}\lambda_{S}^{*}=\sum_{S|e\notin S}\lambda_{S}^{*}=\sum_{S|e\notin S}\frac{1}{(1-w)}\lambda_{S}=\frac{1}{(1-w)}\cdot(1-w)=1
\]

Trivially all bundles in the support of $\{\lambda_{S}^{*}S\}$
are of value at least $APS_{j}(\mathcal{M},b_{j})$, proving the claim.
\end{proof}

\begin{definition}
\label{def:$v^t_p$}
    Let $v_p$ be the valuation function of agent $p$ and let $t\geq 0$ be a scalar, then we define $v_p^t$, the valuation $v_p$ truncated at $t$, as follows:
\[
v_p^t(B)\coloneqq\min\{v_p(B),t\}
\]
\end{definition}

Observe that if $v_p$ is submodular, then also $v_p^t$ is submodular.

\begin{claim}
\label{claim:truncation}
     Let $v_p$ be the valuation function of an agent $p$, and set {$t\leq MMS_p$} (respectively {$t\leq APS_p$}). Then the {new }$MMS$ ($APS$) of agent $p$ {is} $t$ if we consider her valuation function to be $v_p^t$ (\Cref{def:$v^t_p$}). {In the special case of $t=MMS_p$ ($t=APS_p$), this implies that the MMS (APS) of the agent remains unchanged.}
\end{claim}

\begin{proof}
    The partition (fractional partition) that certifies that the MMS (APS) with respect to $v_p$ is at least $t$, certifies the same with respect to $v_p^t$ (because every bundle that has value at least $t$ with respect to $v_p$ has value $t$ with respect to $v_p^t$). The MMS (APS) with respect to $v_p^t$ cannot be larger than $t$, as no bundle has $v_p^t$ value larger than $t$.
\end{proof}


\subsection{{Approximate} APS-fair allocations for submodular agents}
\label{sec:APS}




We now describe an allocation game (introduced in~\cite{BEF21}), that we refer to as the bidding game.
Initially, every agent $i$ is {\em active}, is given a budget of $b^0_i = b_i$ (in particular, in the equal entitlement case $b^0_i = \frac{1}{n}$), and has an empty bundle $S^0_i$ of items. The set of initially unallocated items is denoted by $M^0$. 

The game proceeds in rounds, and in every round, one item is allocated. In round $r \ge 1$, to decide which item is allocated, we do the following.

\begin{enumerate}

\item If there are no active agents, end the allocation algorithm. (The remaining items, if there are any, can be allocated arbitrarily.)

    
    \item {Every active agent $i$ submits a nonnegative bid $p_i^r$ of her choice, not exceeding her budget. Namely, $0 \le p_i^r \le b_i^{r-1}$.}

    \item  The agent $i$ with the highest bid (breaking ties arbitrarily) wins, {and selects an arbitrary item of her choice. Denote this selected item as $e^r$.}  We update $M^r = M^{r-1} \setminus \{e^r\}$ and $S_i^r = S_i^{r-1} \cup \{e^r\}$. Her budget is updated to $b^r_i = b^{r-1}_i - p_i^r$. If $b^r_i =0$, then agent $i$ stops being active. In any case, for agents $j \not= i$, we have $S_j^r = S_j^{r-1}$ and $b^r_j = b^{r-1}_j$.

\end{enumerate}


To help illustrate the key methods used in the proof of \Cref{thm:1/3_APS_guarantee}, we first present a sketch of proof for a weaker version. This will pave the way for the subsequent proof of \Cref{thm:1/3_APS_guarantee}.

\begin{proposition}
\label{pro:third}
    Consider the bidding game described above and an agent $p$ with a submodular valuation function in an equal entitlements setting. Employing a bidding-strategy referred to as $proportional(\frac{1}{3})$ guarantees agent $p$ a minimum value of $\frac{1}{3}\cdot MMS_p$.
\end{proposition}

\begin{proof}
In the bidding game, each agent is initially assigned a budget equal to her entitlement. In the equal entitlement case, each agent receives a budget of $\frac{1}{n}$. The $proportional(\frac{1}{3})$ bidding strategy involves the following steps. Initially, agent $p$ calculates $MMS_p$ (her $MMS$ value).  At the beginning of round $r$, let $\items^r$ denote the set of items that are still available, let $C^r$ denote the set of items that agent $p$ won prior to round $r$, and let $b_p^r$ denote the budget that the agent still holds. In round $r$, agent $p$ bids $\frac{3}{2}\cdot\frac{1}{n}\cdot\frac{1}{MMS_p}\cdot\max_{e\in\items^r}[v_p(e\mid C^r)]$. In other words, the bid of the agent is equal to $\frac{3}{2}$ times the {\em scaled value} of the marginal value of the item of highest marginal value that still remains, where the scaling factor $\frac{1}{n}\cdot\frac{1}{MMS_p}$ is such that after this scaling, the MMS value of $p$ equals the original budget of $p$.
If this bid value exceeds $b_p^r$ (the remaining budget of $p$), then $p$ bids $b_p^r$. In any case, if $p$ wins the bid, she selects the item of highest marginal value in $\items^r$, and pays her bid. We note here that the factor $\frac{3}{2}$ was chosen so as to equal $\frac{1}{2\rho}$, for our choice of $\rho = \frac{1}{3}$. The same type of expression, $\frac{1}{2\rho}$, will appear also in the proof of Theorem~\ref{thm:1/3_APS_guarantee}.

{We now provide a sketch of proof that the above bidding strategy guarantees agent $p$ a bundle of value at least $\frac{1}{3} MMS_p$. In this sketch, $\{B_i\}_{i=1}^n$ denotes an MMS partition for agent $p$. Namely, $v_p(B_i) \ge MMS_p$ for every $i$.}

{Call an item $e$ {\em large} if $v_p(e)>\frac{2}{3}MMS_p$. We claim that, without loss of generality, we can assume that there are no large items. As long as a large item exists, $p$ bids her entire budget. If agent $p$ wins the round, {we are done}. If a different agent $q$ wins the round, that agent spends her entire budget and leaves the bidding game after winning only a single item. 
Intuitively, $q$ did not ``hurt" $p$, since $n-1$ bundles of the MMS partition of $p$ still contain all their items, whereas only $n-1$ agents remain to compete on them. Further details of this argument are omitted.}

According to the bidding strategy, if agent $p$ manages to spend at least half of her budget during the bidding game, she will receive a $\frac{1}{3}$-fraction of her MMS. Therefore, our goal is to show that $p$ manages to spend at least half of her budget.

The main idea is that until $p$ spent half of her budget, other agents cannot do much damage to $p$. The bidding strategy of $p$ has two key properties that are easy to verify. First, the bidding sequence is non-increasing (this is a consequence of submodularity of $v_p$). Second, {in the absence of large items (an assumption that we can make without loss of generality)}, as long as $p$ has spent at most half of her budget, her remaining budget does not constrain her from providing a full bid according to the bidding strategy. Due to this latter property, if another agent $q$ wins an item, {$q$ pays for the item that she takes {at least $\frac{3}{2}$ times} the scaled marginal value that $p$ has for the item.}

Next, we analyze how much harm the other agents {can cause $p$ up to the point when she spends} half of her budget. We denote by $C$ the bundle that $p$ holds {at the last point in time in which she has not spent half her budget.} 
A sufficient condition for $p$ to exceed $\frac{1}{3}MMS_p$ is if there exists a bundle $B_i$ from $p$'s MMS partition that has sufficiently high marginal value (relative to $C$, and after excluding from $B_i$ those items won by other agents) so that together with $C$, the value exceeds $\frac{1}{3}MMS_p$. 
For every item $e$ that another agent wins, that agent pays at least $\frac{3}{2}\frac{1}{n}\frac{1}{MMS_p}v_p(e\mid C)$. {(The fact that we can compare to marginal value relative to $C$ is a consequence of submodularity of $v_p$.)} Hence, the ratio between the value taken from bundles of the MMS partition, and the payment done by the other agents is $\alpha=\frac{3}{2}\frac{1}{n}\frac{1}{MMS_p}$.
Therefore, using the fact that the total budget of all the agents together is 1, we obtain an upper bound on the total value taken by the other agents, which is $\frac{2}{3}\cdot n\cdot MMS_p$. Hence, there must exist a bundle $B_i$ in which the other agents took items of marginal value relative to $C$ of at most $\frac{2}{3}MMS_p$. Hence, together with $C$, there is enough value left in $B_i$ for $p$ to surpass $\frac{1}{3}MMS_p$. (This last argument again uses submodularity of $v_p$.)
\end{proof}

Before presenting the proof of \Cref{thm:1/3_APS_guarantee}, we discuss some of the challenges we will face when extending the (sketch of) proof of Proposition~\ref{pro:third} to the more general setting of \Cref{thm:1/3_APS_guarantee}.

\begin{itemize}
    \item \Cref{pro:third} considers the MMS, whereas \Cref{thm:1/3_APS_guarantee} considers the APS, which is always at least as large as the MMS, and sometimes larger. The analysis needs to be extended to hold relative to this stronger notion, and in particular, can no longer assume the existence of an integral MMS partition.
    \item The setting considered in \Cref{thm:1/3_APS_guarantee} allows for arbitrary entitlements. The proof of \Cref{pro:third} uses the assumption that the setting is that of equal entitlement (for example, in its treatment of large items). 
    \item \Cref{thm:1/3_APS_guarantee} provides a guarantee that is somewhat better than $\frac{1}{3}$-fraction of the APS (that becomes significant if the entitlement is large).
\end{itemize}

We now proceed to present the proof of \Cref{thm:1/3_APS_guarantee}.



{We describe a bidding strategy for the bidding game in the arbitrary entitlement case. It has a parameter $\rho > 0$,
and we refer to it as the proportional bidding strategy $proportional(\rho)$.}  
As we shall later prove, if $\rho$ is chosen to satisfy $\rho \le \frac{1}{3-2b_p}$, then the $proportional(\rho)$ bidding strategy will guarantee that agent $p$ receives a bundle of value at least $\rho APS_p$.
A player $p$ (with valuation function $v_p$ and entitlement $b_p$) 
that uses $proportional(\rho)$ first computes her $APS(\items, v_p, b_p)$  value, which we refer to as $APS_p$. 
Up to some minor technical details (that manifest themselves only if other agents spend their budgets at a rate that is higher than that dictated by the bids of $p$ -- these details do not affect the guarantees offered by the bidding strategy), $proportional(\rho)$ is equivalent to the following simple strategy. Scale the valuation function of $p$ such that her $APS$ equals her entitlement (and budget) $b_p$. In each round, bid $\frac{1}{2\rho}$ times the marginal value (with respect to the items that $p$ already holds) of the item of highest marginal value that is not yet allocated, if $p$ has sufficient budget to do so, and bid the total remaining budget otherwise. If $p$ wins the bid, she selects the item of highest marginal value.

{To simplify the proofs that follow later, we now present the $proportional(\rho)$ bidding strategy in more detail, and introduce terminology that will be used in the proofs.}
We first present $proportional(\rho)$ in the special case in which $v_p(e)\leq 2\rho APS_p$ holds for all items. We refer to the strategy in this special case as $proportional_1(\rho)$. {At the beginning of round $r$, let $\items^r$ denote the set of items not yet allocated, let $C^r$ denote the set of items already allocated to $p$, and let $b_p^{r-1}$ denote the budget remaining for agent $p$. Then the agent bids $\frac{1}{2\rho}\cdot\frac{b_p}{APS_p}\cdot\max_{e\in\items^r}[v_p(e\mid C^r)]$ (the highest marginal value that a yet unallocated item has, scaled by $\frac{1}{2\rho}\frac{b_p}{APS_p}$) if this bid is not larger than $b_p^{r-1}$, and
bids her remaining budget $b_p^{r-1}$ otherwise. If the agent
wins her bid, she selects the item with the highest marginal value with respect to $C^r$.}

We now present the strategy for the remaining case, that in which there are large items that satisfy $v_p(e)>2\rho APS_p$. We refer to the strategy in this case as $proportional_2(\rho)$.

Prior to the beginning of the bidding game, agent $p$ truncates her valuation function, so that the value of each bundle $S$ is $\min[v_p(S), APS_p]$. In the notation of \Cref{def:$v^t_p$}, this new valuation function is denoted as $v_p^t$, with $t=APS_p$. This does not affect $APS_p$, {and preserves submodularity}. To keep notation simple, we still use the notation $v_p$ for this new valuation function.

In the bidding game itself, as long as there exists a large item that satisfies $v_p(e) > 2\rho APS_p$, $p$ bids her entire budget. If the agent
wins her bid, she selects the item with the highest value (which is at least $2\rho APS_p$) and leaves the game (as her budget is exhausted). If the agent does not win any of the large items, let $s$ denote the last round in which there are large items, i.e., from round $s+1$ onward, all unallocated items satisfy $v_p(e) \leq 2\rho APS_p$. At this point, agent $p$ basically switches to using $proportional_1(\rho)$. Below we introduce terminology that describes how this switch is done. This terminology will later be used in our proofs.

Let $I_s$ denote the {\em residual instance}  that remains after round $s$. It includes the set of items that were not yet allocated (which we denote by $\mathcal{\hat{M}}$), 
and each agent has whatever remains from her budget after the $s$ rounds.
View $I_{s}$ as a new allocation instance, that we refer to as $\hat{I_{s}}$. The agents of $\hat{I_{s}}$ are the remaining active agents of $I_{s}$. The set of items of $\hat{I_{s}}$ is $\mathcal{\hat{M}}$ (those items remaining
in $I_{s}$). Setting $\gamma=b^s$ to be the total remaining budget of agents, we update the entitlement of each agent to be $\hat{b_i}=\frac{1}{\gamma} b_i$. Agent $p$ simulates $proportional_1(\rho)$ on $\hat{I_s}$. To do so, for every agent $i \not=p$, a bid $x_{i}^{r}$ in $I_{s}$ is viewed as a bid of $\frac{1}{\gamma}\hat{x}_{i}^{r-s}$ in 
    $\hat{I_{s}}$, and any intended bid $x^r$ of agent $p$ in $\hat{I_{s}}$ is translated to a bid $\gamma x^{r+s}$ in $I_s$. 
(Since the ratio between a budget of an agent in $I_{s}$ and in $\hat{I_{s}}$ is $\frac{1}{\gamma}$, by taking a bid of another agent $i$ in $I_{s}$ and scaling it by $\frac{1}{\gamma}$ we get a legal bid of the agent in $\hat{I_{s}}$. Similarly, scaling by $\gamma$ a bid of agent $p$ in $\hat{I_{s}}$ induces a legal bid in $I_{s}$.)

Observe that by \Cref{claim:APS not decrease in I_s}, the APS of $p$ in $\hat{I_s}$ remains the same as the original value of $APS_p$.

\begin{observation}
\label{obsrv: bidding sequence decreasing}
The sequence of bids of an agent that uses the proportional bidding strategy is weakly decreasing.
\end{observation}

Consider an APS fractional partition 
{$\{\lambda_S\}_{S\subseteq\mathcal{M}}$ for agent $p$} , where {$\sum_S \lambda_S = 1$, $\sum_{S \; | \; e\in S} \lambda_S \le b_p$} 
for every item $e$, and {$v_p(S) \ge APS_p$ for every $S$ with $\lambda_S$ in the support}.
We trace three parameters throughout the execution of the algorithm. One is a {\em lower bound} on the total marginal value of the set of all remaining items to agent {$p$}, 
given her set of items at the time. Initially it is {$L^0 = \sum_S \lambda_S v_p(S) \ge {APS_p}$}. 
Another is the budget of agent $i$, which initially is $b_i^0 = b_i$.  Another is the total remaining budget of all agents. Initially it is $b^0 = 1$.


Consider an agent $p$ with valuation function $v_{p}$. 
We begin by proving \Cref{thm:1/3_APS_guarantee} under the simplifying assumption {that there are no large items i.e., every item satisfies $v_{p}(e) \le 2\rho APS_p$}. 

\begin{lemma}
\label{lem:1/3 guarantee no large items}
For an agent $p$ with a submodular valuation function, if $v_{p}(e) \le 2\rho {APS_p}$ for every item $e$, then by setting $\rho={\frac{1}{3-2b_p}> \frac{1}{3}}$, the $proportional(\rho)$ bidding strategy guarantees agent $p$ a value of at least $\rho \cdot APS_p$.
\end{lemma}

We shall present a sequence of claims that proves \Cref{lem:1/3 guarantee no large items}.

\begin{claim}
\label{claim:2alternatives}
In every round $t$ of the algorithm, at least one of
the following two conditions hold:
\begin{enumerate}
\item Agent $p$'s bid is equal to {$\frac{1}{2\rho}{\cdot\frac{b_p}{APS_p}}\cdot\max_{e \in \items^r}[v_p(e  \mid C^r)]$ (the highest marginal value that a yet unallocated item has, scaled by $\frac{1}{2\rho}{\frac{b_p}{APS_p}}$)} 
\item Agent $p$ already won a bundle with value at least $\rho{APS_p}$.
\end{enumerate}
\end{claim}

\begin{proof}
Suppose that the second condition does not hold. If $p$ did not
win an item yet, then she still has her entire budget, and using the
assumption that no item has a value greater than $2\rho {APS_p}$, condition~1 holds. Otherwise, agent $p$ won items with a total value less than $\rho{APS_p}$ by time $t$. Submodularity of $v_{p}$ implies that the sequence of remaining maximal marginal values  $\max_{e \in \items^r}[v_p(e \; \mid C^r)]$ is
non-increasing in $r$. 
Hence {$\max_{e \in \items^t}[v_p(e \; \mid C^t)] \le \rho APS_p$. Therefore } $\frac{1}{2\rho}{\frac{b_p}{APS_p}}\max_{e \in \items^t}[v_p(e \; \mid C^t)] \le{\frac{1}{2\rho}\rho b_p\leq\frac{1}{2}b_p}\le {b_p^t}$, and condition~1 holds.
\end{proof}

Let $\{\lambda_{S}\}_{S\subseteq\mathcal{M}}$
be the set of weights associated with the APS {for $v_p$} (i.e., for every $S\subseteq\mathcal{M},$
$\lambda_{S}>0\implies v_p(S)\geq {APS_p}$, also $\sum_{S}\lambda_{S}=1$,
and $\forall e\in\mathcal{M},$ $\sum_{S|e\in S}\lambda_{S}\leq b_p$). 

Let $L^{0}\coloneqq\sum_{S}\lambda_{S}v(S)$. Observe that by the
definition of $APS$, $L^{0}\geq APS_{p}$. At the beginning of the
algorithm (when no item has been allocated yet), $L^{0}$ is a lower
bound on the marginal value {that agent $p$ has for} the set of all items. 

Let $f$ denote 
the earliest round after which either all other agents become inactive, or all items have been allocated. Let
$C\subseteq\mathcal{M}$ denote the set of items agent $p$ has by the
end of round $f$, and let $O$ denote the set of items that the other
agents have by the end of round $f$. Define $L^{f}=\sum_{S}\lambda_{S}\cdot v_{p}(S\setminus O\dot{\cup}C\mid C)$.
Namely, $L^{f}$ is the expected marginal value to agent $p$ (who
already holds the set $C$ of items) of a bundle $S$ selected at
random according to the probability distribution over bundles implied
by the coefficients $\lambda_{S}$, after one removes from $S$ those
items that were allocated by the end of round $f$.

\begin{claim}
\label{Claim_first_Bound_Lf}
Let $\tilde{\mathcal{M}}$ be the set of items that remain unallocated after round $f$, i.e., $\tilde{\mathcal{M}}=\mathcal{M}\setminus(O\dot{\cup}C)$. Then $v_{p}(C\cup\tilde{\mathcal{M}})=v_{p}(\mathcal{M}\setminus O)\geq v_{p}(C)+L^{f}$.
In other words, $v_{p}(C)+L^{f}$ is a lower bound on the total
value that agent $p$ will have, if she receives all the remaining items
($\tilde{\mathcal{M}}$).
\end{claim}

\begin{proof}
{If no items remain after round $f$ (i.e., $C\dot{\cup}O=\mathcal{M}$), then} for
each $S\subseteq\mathcal{M}$, $S\setminus O\dot{\cup}C=\emptyset$
and $L^{f}=0$. Hence, $v_{p}(C\cup\tilde{\mathcal{M}}) =  v_p(C) = v_{p}(C)+L^{f}$, proving the claim.

{If items do remain after round $f$, then} every term $v_{p}(S\setminus O\dot{\cup}C\mid C)$
in the sum of $L^{f}=\sum_{S}\lambda_{S}\cdot v_{p}(S\setminus O\dot{\cup}C\mid C)$,
is a marginal value of a partial set of the not-yet-allocated items
(i.e., $(S\setminus O\dot{\cup}C)\subseteq\tilde{\mathcal{M}).}$
Hence $v_{p}(S\setminus O\dot{\cup}C\mid C)\leq v_{p}(\tilde{\mathcal{M}}\mid C)$.
Since the scalars $\{\lambda_{S}\}$ in the sum $L^{f}=\sum_{S}\lambda_{S}\cdot v_{p}(S\setminus O\dot{\cup}C\mid C)$
are non-negative and add up to $1$, we obtain:
\begin{align*}
L^{f}= & \sum_{S}\lambda_{S}\cdot v_{p}(S\setminus O\dot{\cup}C\mid C)\\
\leq & \sum_{S}\lambda_{S}\cdot v_{p}(\tilde{\mathcal{M}}\mid C)\\
= & v_{p}(\tilde{\mathcal{M}}\mid C)
\end{align*}
Hence
\begin{align*}
v_{p}(C)+L^{f}\leq & v_{p}(C)+v_{p}(\tilde{\mathcal{M}}\mid C)=v_{p}(C\cup\tilde{\mathcal{M}})
\end{align*}
\end{proof}

\begin{claim}
\label{Claim_second_bound_Lf}
$\min\{L^{\text{f}}+v_{p}(C),2\rho{APS_p}\}$ is a lower bound on the final
total value of agent $p$.
\end{claim}
\begin{proof}
First, notice that if $p$ is not active in time $f$, then $p$ spent her entire budget, ${b_p}$. The bidding strategy of $p$ (and submodularity of $v_{p})$ implies that in that case,
$p$ has a value of at least $2\rho {APS_p}$ in time $f$, i.e., $v_{p}(C)\geq 2\rho {APS_p}$
and the claim follows in this case.

Otherwise, $p$ is an active agent {after round} $f$. We consider two cases.
If some items remain after round $f$, then agent $p$ is the only remaining active agent. Hence $p$ is the only agent to win items {from $\tilde{\mathcal{M}}$.}
Then, the agent will keep winning items until she becomes inactive or until she wins all remaining items. By \Cref{Claim_first_Bound_Lf}, $v_{p}(\mathcal{M}\setminus O)\geq v_{p}(C)+L^{f}$. The fact that the agent bids at most {$\frac{1}{2\rho}{\cdot\frac{b_p}{APS_p}}$ times} the marginal value of the item she wins in each round guarantees that agent $p$ gets at least $\min\{L^{f}+v_{p}(C), 2\rho {APS_p}\}$. It remains to handle the case of agent $p$ being active at time $f$ while no items remain. In this case, $L^{f}=0$, so the bound is trivial.
\end{proof}

\begin{claim}
\label{claim_L0_Lf}
The following holds:
\[
L^{0}\leq L^{f}+v_{p}(C)+b_p\cdot\sum_{e\in O}v_{p}(e\mid C)
\]
\end{claim}

\begin{proof}
$\!$
\[
L^{f}=\sum_{S}\lambda_{S}\cdot v_{p}(S\setminus O\dot{\cup}C\mid C)=\sum_{S}\lambda_{S}\cdot v_{p}(S\setminus O\mid C)
\]
For every $S\subseteq\mathcal{M}$ we claim:
\[
v_{p}(S)\underset{1.}{\leq}v_{p}(S\mid C)+v_{p}(C)\underset{2.}{\leq}v_{p}(S\setminus O\mid C)+v_{p}(C)+\sum_{e\in S\cap O}v_{p}(e\mid C)
\]
\emph{proof of inequality 1:}
\[
v_{p}(S\mid C)+v_{p}(C)=v_{p}(S\cup C)-v_{p}(C)+v_{p}(C)=v_{p}(S\cup C)\ge v_{p}(S)
\]
\emph{proof of inequality 2:}

Consider an arbitrary order of the set $S\cap O=\{e_{1},\dots,e_{k}\}$.
Then:

\begin{align*}
v_{p}(S\mid C) 
& =v_{p}\left(S\setminus O\mid C\right)+\sum_{i=1}^{k}v_{p}\left(e_{i}\mid\left(S\setminus O\right)\cup C\cup\left(\bigcup_{j=1}^{i-1}e_{j}\right)\right)\\
& \leq v_{p}\left(S\setminus O\mid C\right)+\sum_{i=1}^{k}v_{p}\left(e_{i}\mid C\right)
\end{align*}

Thus, using the last inequality, we obtain the following:

\begin{align*}
L^{0}=\sum_{S\subseteq\mathcal{M}}\lambda_{S}v_{p}(S)
& \leq\sum_{S\subseteq\mathcal{M}}\lambda_{S}\left(
v_p(S\mid C)+v_p(C)
\right) \\
& \leq\sum_{S\subseteq\mathcal{M}}\lambda_{S}\left(v_{p}(S\setminus O\mid C)+v_{p}(C)+\sum_{e\in S\cap O}v_{p}(e\mid C)\right) \\
& \underset{*}{\leq}v_{p}(C)+b_p\cdot\sum_{e\in O}v_{p}(e\mid C)+\sum_{S\subseteq\mathcal{M}}\lambda_{S}v_{p}(S\setminus O\mid C)\\
& =v_{p}(C)+b_p\cdot\sum_{e\in O}v_{p}(e\mid C)+L^{f}
\end{align*}
where inequality * is since each item $e\in\mathcal{M}$ has a total
weight of at most $b_p$ (by the APS definition). Overall, as we wanted
to show, we obtained the following:
\[
L^{0}\leq L^{f}+v_{p}(C)+b_p\cdot\sum_{e\in O}v_{p}(e\mid C)
\]
\end{proof}

\begin{claim}\label{claim_other_agent_items_reducing_Lf}
Either $v_p(C) \ge \rho {APS_p}$, or
\[
\sum_{e\in O}v_{p}(e\mid C)\leq {2\rho(b^0-b_p){\cdot \frac{APS_p}{b_p}}}
\]
\end{claim}

\begin{proof}
Suppose that $v_p(C) < \rho {APS_p}$. Let $C^{e}\subseteq C$ be the set of items agent $p$ already won
when another agent wins item $e$.
\Cref{claim:2alternatives} implies that
agent $p$ bids {$\frac{1}{2\rho}{\cdot\frac{b_p}{APS_p}}$ times }the highest marginal value of an item w.r.t $C^{e}$.
Hence, when the other agent $i$ wins item $e$, agent $p$ bid is
at least ${\frac{1}{2\rho}{\frac{b_p}{APS_p}}}v_p(e\mid C^{e})\geq {\frac{1}{2\rho}{\frac{b_p}{APS_p}}}v_p(e\mid C)$, and winning item $e$ reduces the budget of agent $i$ by at least ${\frac{1}{2\rho}{\frac{b_p}{APS_p}}}v_{p}(e\mid C)$. Since the
budget of all other agents at the beginning is $b^{0}-{b_p}$, we
obtain 
\[
\sum_{e\in O}\frac{b_p}{2\rho APS_p}v_{p}(e\mid C)\leq\sum_{e\in O}\frac{b_p}{2\rho APS_p}v_{p}(e\mid C^{e})\leq {(b^0-b_p)}
\]
The claim follows by rearranging (scaling both sides by $\frac{2\rho APS_p}{b_p}$).
\end{proof}

We are now ready to prove Lemma~\ref{lem:1/3 guarantee no large items}.

\begin{proof}
Considering \Cref{claim_L0_Lf} and \Cref{claim_other_agent_items_reducing_Lf}, we have that either $v_p(C) \ge \rho{b_p}$, or the following holds:
\[
APS_p\leq L^0\leq v_{p}(C)+L^{f}+b_p\cdot\sum_{e\in O}v_{p}(e\mid C) \leq v_{p}(C)+L^{f}+{b_p\cdot2\rho{\frac{APS_p}{b_p}}(b^0-b_p)}
\]

By rearranging the above, and plugging $b^0=1$ we obtain:
\[
 APS_p\cdot {(1-2\rho b^0+2\rho b_p)} = APS_p\cdot {(1-2\rho +2\rho b_p)} 
 \le v_{p}(C)+L^{f}
\]

By setting $\rho=\frac{1}{3-2b_p}$ 
the above gives $L^{f}+v_{p}(C) \ge \rho{APS_p}$. {So far we obtained that either $v_p(C)\geq\rho APS_p$, or } 
agent $p$ is guaranteed to have a total final value of $\min\{L^{f}+v_{p}(C),2\rho {APS_p}\}$ (\Cref{Claim_second_bound_Lf}). 
{ Hence, in both cases, when setting $\rho=\frac{1}{3-2b_p}$, agent }$p$ is guaranteed to have at least $\rho$ fraction of her $APS$. In the special case of equal entitlements (where $b_p=\frac{1}{n}$) it implies $\rho=\frac{n}{3n-2}$. This completes the proof of ~\Cref{lem:1/3 guarantee no large items}.

\end{proof}

We now restate and prove \Cref{thm:1/3_APS_guarantee}.
\APSbidding*
\begin{proof}
~\Cref{lem:1/3 guarantee no large items} handles the case that $v_p(e) \le 2\rho {APS_p}$ for every item $e$.
It remains to handle the case that there are items $e$ of value $v_p(e) > 2\rho {APS_p}$.

Consider an input instance $I_0$. 
Run the bidding game with $p$ using the proportional bidding strategy. As described in $proportional_2$, {$s$ denotes the {last round} in which there was an unallocated item $e$ with $v_p(e) > 2\rho APS_p$.} 
If agent $p$ won some item by the end of round $s$, then she has a value of at least $2\rho APS_p$, and we are done. Hence, we may assume that agent $p$ did not win any item in the first $s$ rounds. {Recall the definitions of residual instance $I_s$, $\hat{I_s}$ and $\gamma$ (which is the scaling factor between bids in $I_s$ and $\hat{I_s}$) presented in $proportional_2$.}

By the definition of $s$, in each of the first $s$ rounds,
there is an available item with $v_{p}(e)>2\rho APS_p$. Thus agent {$p$ bids her entire budget $b_p$ in each such round}, and the (other) agent who wins the round spends at least $b_p$. (In the special case of equal entitlement, this means that in each of the first $s$ rounds, some agent wins a single item and becomes inactive. Hence, in the residual instance $I_s$ there are $n-s$ active agents (including $p$), each active agent has her entire original budget, and no active agent has any items.)

Setting $\gamma=b^s\leq 1 - s\cdot b_p$, the entitlement of each agent {in $\hat{I_s}$} is $\hat{b_i^s}=\frac{1}{\gamma} b_i^s$. 


{
Recall that agent $p$ simulates the bidding strategy ($proportional_1$) on $\hat{I_{s}}$. A bid $p_{i}^{r}$ in $I_{s}$ is interpreted as a bid of $\frac{1}{\gamma}\hat{p}_{i}^{r-s}$ in $\hat{I_{s}}$. 
By \Cref{claim:APS not decrease in I_s}, the APS of agent $p$ at $\hat{I_{s}}$ 
{stays the same as the} APS of $p$ in the original instance. Hence, in $I_s$, agent $p$ will get the same bundle as she gets in the run on $\hat{I_{s}}$. By \Cref{lem:1/3 guarantee no large items}, this bundle is of value at least $\rho APS_p$, as desired.}
\end{proof}

\subsection{Approximate MMS-fair allocations for submodular agents}
\label{sec:Approximated MMS-fair existance}

In the standard version of the bidding game, every agent has an initial budget, and in each round, the highest bidder picks an item. Each agent plays until she exhausts her budget, and the game ends when either there are no more items left, or all agents exhaust their budgets. Now we define another version of the bidding game, in which agents might leave the game before exhausting their budget.

    
\begin{definition}
    The  $\rho$-altruistic version of the bidding game is the same bidding game with the change that every agent becomes inactive after spending a $\rho$-fraction of her budget. 
\end{definition}


We shall consider a bidding strategy for the $\rho$-altruistic bidding game, that we shall refer to as the {\em proportional bidding strategy}. We make two assumptions that simplify our presentation. These assumptions have no effect on the correctness of Theorem~\ref{thm:equal-10/27} that follows. These assumptions concern the submodular valuation function $v_p$ of agent $p$ that uses the proportional bidding strategy.

\begin{enumerate}
    \item The valuation function is scaled so that $MMS_p = b_p$ (the MMS of agent $p$ equals her entitlement).
    \item The valuation function is truncated at $MMS_p$ (as in Definition~\ref{def:$v^t_p$}). By Claim~\ref{claim:truncation}, this truncation does not affect the MMS value. 
\end{enumerate}

We now present the proportional bidding strategy, as used by agent $p$. At the beginning of round $r$, let $\items^r$ denote the set of items not yet allocated, let $C^r$ denote the set of items already allocated to $p$, and let $b_p^r$ denote the budget remaining for agent $p$. Then the agent bids $\max_{e\in\items^r}[v_p(e\mid C^r)]$  (the highest marginal value that a yet unallocated item has) 
if this bid is not larger than $b_p^{r}$, and bids her remaining budget $b_p^{r}$ otherwise. 

Recall \Cref{thm:equal-10/27}:
\altruistic*


Before presenting the proof of Theorem~\ref{thm:equal-10/27}, we sketch the proof for a weaker version of the theorem, setting $\rho = \frac{4}{11}$ instead of $\rho = \frac{10}{27} > \frac{4}{11}$. Already this weaker version improves over the ratio of $\rho = \frac{n}{3n-2}$ of Theorem~\ref{thm:1/3_APS_guarantee} (when $n > 8$). Moreover, the proof of this weaker version of Theorem~\ref{thm:equal-10/27} conveys some intuition that may be helpful for following the proof of Theorem~\ref{thm:equal-10/27}. 


\begin{proposition}
\label{pro:4/11}
In the equal entitlement case with submodular valuations, for $\rho = \frac{4}{11}$, every agent that uses the proportional strategy in the altruistic version of the bidding game is guaranteed to get a bundle of value at least a $\rho = \frac{4}{11}$ fraction of her MMS. 
\end{proposition}

\begin{proof}
We only sketch the proof, as we shall later present a full proof for Theorem~\ref{thm:equal-10/27}.

If agent $p$ that uses the proportional bidding strategy manages to spend $\rho \cdot b_p$, then she also wins items of total value at least $\rho \cdot MMS_p$, and we are done. Hence it remains to exclude the case that agent $p$ failed to spend $\rho b_p$. In this case,  partition the agents other than $p$ into three classes, $X_0, X_1, Y$. 

Class $X_0$ contains those agents that by the end of the bidding game take only one item. 
Intuitively, an agent $i$ of class $X_0$ who took one item does not ``hurt" $p$, because there are $n-1$ bundles in the MMS partition of $p$ from which agent $i$ does not take any item, and only $n-1$ agents (including $p$) compete for items in these bundles. Hence we may assume that no agent is in class $X_0$. See further details in \Cref{claim: no X_0 agents}. 

Class $X_1$ contains those agents that by the end of the bidding game take two items. Being in the $\rho$-altruistic bidding game implies that for the first item that such an agent $i$ took she paid at most $\rho b_p$ (here we use the fact that agents have equal entitlements), implying that the bid of $p$ at the time was at most $\rho b_p = \rho MMS_p$. By the proportional strategy, this bid was equal to the highest marginal value for $p$ for any of the remaining items at the time, and as the sequence of highest marginal values cannot increase as rounds progress, this further implies that the marginal $v_p$ values of items taken by agent $i$ is at most $2 \rho MMS_p$. A key observation is that if there are more than $\frac{n}{2}$ agents in class $X_1$, then there must be a bundle $B$ in the MMS partition of $p$ from which they took at least two items. As it does not matter for $p$ which of the agents in $X_1$ takes which of the items that the agents in $X_1$ collectively take (as long as each agent in $X_1$ takes two items, and each such item has marginal value at most $\rho b_p$), we may pretend that there is an agent $i$ in $X_1$ for which the two items that she takes are from this bundle $B$. This agent $i$ does not ``hurt" $p$, because there are $n-1$ bundles in the MMS partition of $p$ from which agent $i$ does not take any item, and only $n-1$ agents (including $p$) compete for items in these bundles. Hence, we may assume that at most $n/2$ agents are in class $X_1$. See further details in \Cref{claim: at most n/2 X_1  agents}.

Class $Y$ contains all remaining agents. Such an agent $i$ either takes no items (and then of course she does not hurt $p$), or takes only one item with marginal value (to $p$) of at most $\rho APS_p$, or takes $k \ge 3$ items. In the latter case, being in the $\rho$-altruistic bidding game implies that for her first $k-1$ items agent $i$ spent at most $\rho b_p$, and hence the bid of $p$ for the last item taken by $i$ was at most $\frac{1}{k-1} \rho b_p$. This implies that the total marginal $v_p$ values (with respect to items held by $p$) of items taken by $i$ is at most $\frac{k}{k-1} \rho MMS_p$. For $k \ge 3$, this is maximized when $k=3$, giving $\frac{3}{2} \rho MMS_p$.

Summing up, agents other than $p$ take a total marginal $v_p$ value of at most $\frac{n}{2}\cdot 2\rho MMS_p + (\frac{n}{2}-1)\cdot \frac{3}{2} \rho MMS_p = (\frac{7n}{4} - \frac{3}{2}) \rho MMS_p$. Hence from at least one of the bundles $B$ of the MMS partition of $p$, the total marginal value taken by other agents is at most $(\frac{7}{4} - \frac{3}{2n}) \rho MMS_p$. As the bidding game ended with $p$ spending strictly less than $\rho b_p = \rho MMS_p$, no items left in $B$ have any marginal value for $p$. Denoting the bundle that $p$ receives by $C$, this implies that $v_p(B\mid C)\leq (\frac{7}{4} - \frac{3}{2n})\rho MMS_p$. Hence. we have that $MMS_p \le v_p(B) \le v_p(B\mid C) + v_p(C) \le (\frac{7}{4} - \frac{3}{2n})\rho MMS_p + v_p(C)$. For $\rho = \frac{4}{11}$, the assumption that $v_p(C) < \rho MMS_p$ leads to a contradiction in the above inequality, thus proving that $v_p(C) \ge \rho MMS_p$, as desired.
\end{proof}

Observe that due to the slackness factor of $\frac{3}{2n}$ in the last paragraph of the proof of Proposition~\ref{pro:4/11}, we can adapt the proof to get a slightly better bound for $\rho$, of the order of $\frac{4}{11} + \Theta(\frac{1}{n})$. Though this slackness term does not substantially change the approximation ratio, it does prove useful for designing a polynomial time algorithm that outputs an allocation in which each agent receives at least a $\frac{4}{11}$ fraction of her APS. See more details in Section~\ref{sec:polynomial}.

Having seen the proof of Proposition~\ref{pro:4/11}, let us explain the source of improvement that leads to the proof of Theorem~\ref{thm:equal-10/27}. The $\frac{4}{11}$ approximation ratio (rather than a better one) comes from the possibility that agents other than $p$ take $2 \cdot \frac{n}{2}$ items of value $\rho MMS_p$, and $3 \cdot (\frac{n}{2}-1)$ items of value $\frac{1}{2}\rho MMS_p$, for a total value of nearly $\frac{7}{4} \rho MMS_p$. However, in this case the other agents take fewer than $3n$ items, implying that in at least one of the bundles of the MMS partition of $p$, they take at most two items, of values $\rho MMS_p$ and ${\frac{1}{2}\rho} MMS_p$ (recall that we may assume that no two items of value $\rho MMS_p$ are in the same bundle of $p$'s MMS partition). Hence one of these bundles still has value of $MMS_p - \frac{3}{2}{\rho}MMS_p$. The assumption that $p$ gets a value of at most $\rho MMS_p$ then implies {that} $MMS_p \le \frac{5}{2}\rho MMS_p$, implying that $\rho \ge \frac{2}{5}$.

The other agents may do damage to $p$ in a different way. $\frac{n}{2}$ agents might each take two items of value $\rho MMS_p$, $\frac{n}{3}$ agents might each take three items of value $\frac{1}{2}\rho MMS_p$, and $\frac{n}{6} - 1$ agents might each take six items of value $\frac{1}{5}\rho MMS_p$. Ignoring the missing one agent (that is important, but is ignored only for the sake of the argument), this allows the other agents to take three items from each MMS bundle, of values $\rho MMS_p$, $\frac{1}{2}\rho MMS_p$ and $\frac{1}{5}\rho MMS_p$. This leads to the inequality $MMS_p \le \frac{27}{10}\rho MMS_p$, implying that $\rho \ge \frac{10}{27}$. The proof of Theorem~\ref{thm:equal-10/27} shows that this is the most damage that the other agents can do. 
\bigskip

We now present rigorous proofs for two claims that were used in the proof of \Cref{pro:4/11}, and will also be used in the proof of \Cref{thm:equal-10/27}. These claims imply that we can assume without loss of generality that there are no agents of type $X_0$, and at most $\frac{n}{2}$ agents of type $X_1$. For both claims, we present their proofs under a framework in which we prove by induction on $n$ (the number of players) the statement that the proportional bidding strategy guarantees a $\rho$ fraction of $MMS_{p}$ to agent $p$. The statement is clearly true for the base case of $n = 1$. Hence for a given value of $n > 1$, we just prove the inductive step (assuming that the statement has already been proved for all $n' < n$). To simplify the presentation of the proofs of the claims, we make two assumptions. We stress that these assumptions do not affect the correctness of the claims. (Alternatively, these assumptions can be turned into facts, by simply incorporating them as part of the description of the proportional bidding strategy.)

\begin{itemize}

\item Agent $p$ truncates her valuation function at $t=MMS_{p}$, (i.e.,
$v_{p}\leftarrow v_{p}^{t}$).

\item Agent $p$ fixes some order over $\mathcal{M}$ in order to break
ties consistently. If she wins a round and there is more than one item with the highest marginal value, she will break the tie by picking the item appearing earlier in this order.

\end{itemize}

\begin{claim}
\label{claim: no X_0 agents}
In the inductive framework presented above, suppose that there is an allocation instance $I$ with $n$ agents, and that agent $p$ has a submodular valuation function $v_p$ and uses the proportional bidding strategy. If in the run of the bidding game there is at least one agent of type $X_{0}$, then agent $p$ is guaranteed to receive a bundle of value at least $\rho MMS_p$.
\end{claim}

\begin{proof}
Recall that by our assumption, the original valuation function $v_p$ is truncated so that no bundle has value larger than $MMS_p$. Consider a run $R$ of the bidding game, in which agent $p$ uses the proportional bidding strategy with an order 
$\Pi$ over the items $\items$.
Suppose that in this run agent $i$ is of type $X_0$. Let $e$ denote the single item that agent $i$ takes, and let $r$ denote the round number in which $i$ took item $e$. Suppose for the sake of contradiction that in this run $R$ agent $p$ receives a bundle of value strictly smaller than $\rho MMS_p$. Then we show a new allocation instance $I'$ with only $n-1$ agents, and a run $R'$ in which an agent $p'$ that uses the proportional bidding strategy gets value smaller than $\rho {MMS_{p'}}$. This contradicts our induction hypothesis. 

The instance $I'$ contains $n-1$ agents and the set $\items' = \items \setminus \{e\}$ of items. The valuation $v_{p'}$ is identical to $v_p$, though defined only over $\items'$ (for every $S \subseteq \items'$ we have that $v_{p'}(S) = v_p(S)$). Note that even though there are $n-1$ agents, $MMS_{p'} = MMS_p$. (The MMS partition for $v_p$, after dropping the bundle containing $e$, can serve as an MMS partition for $v_{p'}$, showing that $MMS_{p'} \ge MMS_p$. Being truncated at $MMS_p$, we have that $MMS_{p'} \le MMS_p$.) The order {$\Pi'$} used by $p'$ over $\items'$ is identical to {$\Pi$} (but with item $e$ removed). The run $R'$ is identical to $R$, except for the following four changes:
\begin{itemize}
    \item Agent $i$ and her bids are removed.
    \item Agent $p'$ plays in $R'$ the role that agent $p$ played in $R$.
    \item Round $r$ (the one in which item $e$ was taken) is removed.
    \item All bids are scaled by $\frac{n-1}{n}$ (as the budgets of agents are $\frac{1}{n-1}$ rather than $\frac{1}{n}$)
\end{itemize} 
Consequently, the sequence of bids and item choices that $p'$ makes in $R'$, being derived from the bids of $p$ in $R$, is consistent with $p'$ using the proportional strategy. The bundle received by $p'$ in $R'$ is exactly the same bundle that $p$ received in $R$, and hence of value below $\rho {MMS_{p'}}$. This contradicts our induction hypothesis. 
\end{proof}

\begin{claim}
\label{claim: at most n/2 X_1  agents}
In the inductive framework presented above, suppose that there is an allocation instance $I$ with $n$ agents, and that agent $p$ has a submodular valuation function $v_p$ and uses the proportional bidding strategy. If in the run of the bidding game, more than $\frac{n}{2}$ agents are of type $X_{1}$, then agent $p$ is guaranteed to receive a bundle of value at least $\rho MMS_p$.
\end{claim}

\begin{proof}
Consider a run $R$ of the bidding game, in which agent $p$ uses the proportional bidding strategy with an order $\Pi$ over the items $\mathcal{M}$. Suppose that in this run, more than $\frac{n}{2}$ agents are of type $X_1$. Suppose for the sake of contradiction that in this run $R$ agent $p$ receives a bundle of value strictly smaller than $\rho MMS_p$. Then we show a new instance $I'$ with only $n-1$ agents, and a run $R'$ in which agent $p'$ that uses the proportional bidding strategy gets a value strictly smaller than $\rho MMS_{p'}$. This contradicts our induction hypothesis.

Consider an MMS partition $\{B_i\}_{i=1}^n$ with respect to $v_p$. Since there are more than $\frac{n}{2}$ agents of type $X_1$, there are at least $n+1$ items taken by these agents. Hence, there exists a bundle $B_i$ in the MMS partition that contains at least two of these items. Denote these items by $e_1$ and $e_2$, the agents who win them by $a_1$ and $a_2$, and the rounds in $R$ in which these items were taken by $r_1$ and $r_2$. 

The instance $I'$ contains $n-1$ agents and the set $\mathcal{M}'=\mathcal{M}\setminus\{e_1,e_2\}$ of items. The valuation $v_p'$ is identical to $v_p$, though defined only over $\mathcal{M}'$ (for every $S\subseteq\mathcal{M'}$ we have that $v_p'(S)=v_p(S)$). Note that even though there are $n-1$ agents, $MMS_{p'}=MMS_p$. ($\{B_i\}_{i=1}^n$, the MMS partition for $p$, can serve as an MMS partition for $p'$, after dropping the bundle containing $e_1$ and $e_2$, showing that $MMS_{p'}\geq MMS_p$. Being truncated at $MMS_p$, we have that $MMS_{p'}\leq MMS_p$). The order $\Pi'$ used by $p'$ over $\mathcal{M}'$ is identical to $\Pi$ (but with items $e_1,e_2$ removed). The run $R'$ is identical to $R$, except for the following changes:

\begin{itemize}

    \item Rounds $r_1$ and $r_2$ (the rounds in which items $e_1, e_2$ were taken) are removed.
    
    \item Agent $p'$ plays in $R'$ the role that agent $p$ played in $R$.

    \item Agent $a_1$ and her bids are removed. 
    
    \item If $a_1\neq a_2$ (i.e $e_1$ and $e_2$ were taken by different agents) then let $e_3$ denote the other item taken by $a_1$ in $R$, and let $r_3$ denote the round in which it was taken. In $R'$ 
    agent $a_2$ takes item $e_3$ in round $r_3$ (instead of $e_2$ that $a_2$ took in the run $R$ - recall that $e_2 \not\in \items'$).
    
    \item All bids are scaled by $\frac{n-1}{n}$ (as the budgets of agents are $\frac{1}{n-1}$ rather than $\frac{1}{n}$)

\end{itemize}

Consequently, the sequence of bids and item choices that $p'$ makes in $R'$, being derived from the bids of $p$ in $R$, is consistent with $p'$ using the proportional strategy. The bundle received by $p'$ in $R'$ is exactly the same bundle that $p$ received in $R$, and hence of value below $\rho MMS_{p'}$. This contradicts our induction hypothesis.
\end{proof}



{We are now ready to prove \Cref{thm:equal-10/27}.}

\begin{proof}
Consider an arbitrary allocation instance with equal entitlement, and an arbitrary agent $p$ with a submodular valuation function $v_p$. We wish to show that by using the proportional strategy in the $\rho$-altruistic version of the bidding game (with a choice of $\rho = \frac{10}{27}$), $p$ receives a bundle of value at least $\frac{10}{27} \cdot MMS_p$. Equivalently (by scaling $v_p$ so that $MMS_p = \frac{27}{10}$), we wish to show that if $MMS_p = 2.7$, then $p$ receives a bundle of value at least~1. 

We begin by presenting a claim that will be used later.
Let $\left\{ B_{i}\right\} _{i=1}^{n}$ be an MMS partition of agent $p$, i.e., for each $i$ $v_{p}(B_{i})\geq MMS_{p}$. Let $f$ be the earliest round after which no other agents are active. Let $C$ denote the bundle $p$ has by the end of round $f$, and let $O\subseteq\mathcal{M}$ be the set of items taken by other agents.
The following claim is similar in nature to \Cref{Claim_second_bound_Lf}

\begin{claim}
\label{claim: No large value at each MMS bundle}
In the altruistic version of the bidding game, for each $i\in[n]$, the final value that $p$ has is at least 
\[\min\left\{ \rho MMS_{p},v_{p}(C)+v_{p}(B_{i}\setminus O\mid C)\right\} = \min\left\{ \rho MMS_{p},v_{p}(C
\cup\left\{  B_{i}\setminus O\right\})\right\}\]
\end{claim}

\begin{proof}
If $p$ is not active at the end of round $f$, then $p$ has a value of at least $\rho MMS_{p}$ at that time, i.e., $v_{p}(C)\geq\rho APS_{p}$, as desired. 

If $p$ is active at the end of round $f$, we consider two cases. If some items remain after round $f$, then agent $p$ is the only remaining active agent. Hence $p$ is the only agent to win items from $\mathcal{M}$. Then, the agent will keep winning items until either she becomes inactive (and has value at least $\rho MMS_{p}$) or until she wins all remaining items from $B_{i}\setminus O$. It remains to handle the case of agent $p$ being active after time $f$ while no items remain. In this case, $v_{p}(B_{i}\setminus O)=v_{p}(\emptyset)=0$, so the bound is trivial. 
\end{proof}



Every negative example (i.e., an instance of the problem in which agent $p$ does not get at least $\rho MMS_p$) implies (by scaling the valuation function of the agent) the existence of another instance of the problem, in which $MMS_p = \frac{1}{\rho}$ and $p$ gets a bundle of value less than 1. Denote the $MMS$ of agent $p$ by $z$.
We set up a system of linear inequalities. The linear inequalities encode various constraints on the $v_p$ value of items that agents other than $p$ receive, given that $p$ is using the proportional strategy and ends up with a value of at most $1$. We show that the system of inequalities is feasible only if $z \le 2.7$. {This implies that 
$\rho \geq \frac{10}{27}$}. 



To simplify the presentation of the system of inequalities, we shall have a slight abuse of terminology. The term {\em payment} of an agent (say, for an item in round $r$) will correspond to the bid of agent $p$ (in round $r$), even though the actual payment might be larger (if the bid of the winning agent was strictly higher than the bid of agent $p$). With this abuse of notation, the sequence of payments that an agent makes is nonincreasing. This implies that the total payment of an agent that wins $t > 1$ items is at most $\frac{t}{t-1}$, because otherwise the agent spent more than~1 on the first $t-1$ items that she picked, and would become inactive before winning its $t$th item. Observe that the total payment of $p$ can be assumed to be not more than~1, as otherwise, the bundle that $p$ receives according to the proportional strategy has $v_p$ value at least~1, and we are done.

Consider a negative example for some $\rho$, i.e., an instance in which an agent does not get a $\rho$ fraction of her MMS, and she gets a value of at most $1$. Keeping this instance in mind, we will present our set of inequalities, and explain why the instance respects each of them.

The system of inequalities has nonnegative variables, $x_{1}, x_{2}, x_{3}, x_{4}, y, q, z$. For each $1 \le i \le 4$, $x_{i}$ represents the fraction of agents who satisfy the following two conditions:
\begin{enumerate}
\item The agent takes $i+1$ items.
\item The total payments that the agent made is at least $\frac{6}{5}$.
\end{enumerate}

The reason why we do not introduce a variable $x_0$ (for agents who take one item and pay at least $\frac{6}{5}$) is because Claim~\ref{claim: no X_0 agents} implies that we can assume that there are no such agents. This also implies that a payment for an item is never larger than~1.
Variable $y$ represents the fraction of the rest of the agents, those that either take at least six items or paid at most $\frac{6}{5}$. Observe that in either case, any such agent paid a total of at most $\frac{6}{5}$. 
Variable $z$ represents the MMS of agent $p$ (and recall that we scale the valuation $v_p$ so that $MMS_p = \frac{1}{\rho}$).
The variable $q$ needs a more detailed explanation. For every $0\le s\le\frac{1}{2}$, let $\alpha_{s}$ denote the fraction of agents that satisfy the following conditions:

\begin{enumerate}
\item The agent takes at least three items.
\item The total payments made by the agent is larger than $\frac{6}{5}$.
\item Her payment for the first item that she takes is $\frac{1}{2}+s$. 
\end{enumerate}

Note that this implies that the number of items that such an agent takes is either three, in which case her total payments are at most $\frac{3}{2}-s$, or four, in which case her total payments are at most $\frac{5}{4} - \frac{s}{2}$, which is smaller than $\frac{4}{3}-s$ for $s < \frac{1}{6}$ (note that if $s \ge \frac{1}{6}$ then her total payments when taking four items are at most $\frac{6}{5}$, and hence this case is excluded). The variable $q$ represents $\int_{0}^{\frac{1}{2}}s\cdot\alpha_{s}ds$.

{We turn to present the linear inequalities, and explain why each of these inequalities must hold if $p$ executes the proportional bidding strategy}

\begin{enumerate}

\item $x_{1}+x_{2}+x_{3}+x_{4}+y \le 1 - \frac{1}{n}$. The variables $x_1, x_2, x_3, x_4, y$ represent fractions of the total number of agents, and agent $p$ is not included in these fractions. As the total number of agents is $n$, the sum of the fractions is at most $1 - \frac{1}{n}$.

{\item $z-2x_{1}-(3/2)x_{2}-(4/3)x_{3}-(5/4)x_{4}-(6/5)y+q \leq 1$. Fix
$\left\{ B_{i}\right\} _{i=1}^{n}$, an MMS partition for agent $p$.
Since we assume $p$ is active at the end of the algorithm, \Cref{claim: No large value at each MMS bundle} implies that the final value of $p$ is at least $v_{p}(C) + v_{p}\left\{ B_{i}\setminus O\mid C\right\} )$. 
Since the final value of $p$ is at most $1$, we obtain $1\geq v_{p}(C) + v_{p}\left\{ B_{i}\setminus O\mid C\right\} )$

Recall that $O\mathcal{\subseteq M}$ is the set of items taken by other agents. Consider the following partition of $O$ , $\left\{ O_{i}=B_{i}\cap O\right\} _{i=1}^{n}$. Then, for each $i$, 
\begin{align*}
1 & \geq v_{p}(C)+v_{p}\left\{ B_{i}\setminus O\mid C\right\} )\\
 & =v_{p}(C)+v_p(B_{i}\setminus O_{i}\mid C)\\
 & \geq v_{p}(C)+v_{p}(B_{i}\mid C)-v_{p}(O_{i}\mid C)\\
 & =v_{p}(C\cup B_{i})-v_{p}(O_{i}\mid C)\\
 & \geq v_{p}(B_{i})-v_{p}(O_{i}\mid C)\\
 & =z-v_{p}(O_{i}\mid C)
\end{align*}

Denote the total payments of agents other than $p$ as $P_O$. Notice that $\sum_{i}v_{p}(O_{i}\mid C)\leq\sum_{i}\sum_{e\in O_i} v_{p}(e\mid C)\leq P_O$.
 Hence, there exists $i\in[n]$ for which $v_{p}(O_{i}\mid C)\leq\frac{1}{n}\cdot P_O$.
 
We now upper bound $\frac{1}{n}\cdot P_O$. Recall that the total payment of an agent that takes $t > 1$ items is at most $\frac{t}{t-1}$. Moreover, if agents of type $x_{2}$ and $x_{3}$  have a payment strictly larger than $\frac{1}{2}$ for their first item, then they do not reach the maximum payment they can achieve ($\frac{3}{2}$ and $\frac{4}{3}$). By the definition of $q$, we obtain that their total payment is reduced by at least $q$. (Recall the discussion that follows the definition of $q$. It shows that if an agent of type $x_{2}$ pays $\frac{1}{2}+s$ for her first item, then the maximum payment she might reach after taking her three items is at most $\frac{3}{2}-s$. Likewise, it shows that if an agent of type $x_{3}$ pays $\frac{1}{2}+s$ for her first item, then the maximum payment she might reach after taking her four items is at
most $\frac{4}{3}-s$.) Therefore we have
\[
\frac{1}{n}\cdot P_O {\leq}2x_{1}+(3/2)x_{2}+(4/3)x_{3}+(5/4)x_{4}+(6/5)y-q
\]
The constraint follows by using this last inequality and the two previous inequalities $1 \geq z-v_{p}(O_{i}\mid C)$ and $v_{p}(O_{i}\mid C) \le \frac{1}{n}P_0$.
\item $2x_{1} \le 1$. 
We can assume that at most half of the agents are of type $x_1$, by \Cref{claim: at most n/2 X_1  agents}.

{
\item $\frac{6}{5}y+q \geq (z-1-\frac{3}{2})\cdot\left(3-2x_{1}-3x_{2}-4x_{3}-5x_{4}\right)$. {(This is not a linear inequality, but it will be linearized before it will be used.)}
We refer to items taken by agents represented by $x_{1},x_{2},x_{3},x_{4}$ as \emph{primary}, and to items taken by agents represented by $y$ as \emph{secondary}. How much payment do secondary items need to have so that payment of at least $z-1$ is paid for items in each bundle $B_i$ of the MMS partition (which is a necessary condition for $p$ having a total value of at most $1$ \Cref{claim: No large value at each MMS bundle})? The total number of primary items is $(2x_{1}+3x_{2}+4x_{3}+5x_{4})n$. We split the argument into two cases.

In the first case, there are at most $3n$ \emph{primary} items. The number of primary items missing to complete this number to $3n$ is $(3-2x_{1}-3x_{2}-4x_{3}-5x_{4})n$. {We refer to each such missing item as a \emph{deficiency unit}.} As noted above, we assume no two primary items of agents of type $x_{1}$ are taken from a single bundle. 
{We analyze the distribution of \emph{deficiency units} over the MMS bundles when a bundle with $\ell\leq3$ \emph{primary} items has $3-\ell$ \emph{deficiencies}. Then, we will find properties of the distribution of \emph{deficiencies} that minimizes the amount of payment to be paid by $y$ agents to reach $z-1$ payment in each MMS-bundle.

\begin{itemize}


    \item A bundle $B_j$ in the MMS partition with one \emph{deficiency unit}, has a payment for \emph{primary} items of at most $1+\frac{1}{2}+s$.  At least $z-1-\frac{3}{2}-s$ needs to be paid  by $y$ agents in order to surpass a $z-1$ payment in $B_j$. I.e., a $z-1-\frac{3}{2}-s$ for the one unit of deficiency.

    \item A bundle $B_j$ in the MMS partition with two \emph{deficiency units} has a payment for \emph{primary} items of at most $1$. At least $z-1-1$ needs to be paid by $y$ agents to surpass a $z-1$ payment in $B_j$. I.e., a $\frac{z-2}{2}$ on average per \emph{deficiency unit}.

    \item A bundle with three \emph{deficiency units}, needs a payment of at least $z-1$ by $y$ agents. I.e., a $\frac{z-1}{3}$ on average for per deficiency unit.
    \end{itemize}
    
    For $z\leq 3$, the minimum payment per deficiency occurs when every deficiency unit is in a different bundle (i.e., these bundles have one deficiency unit, the case of bullet two).}
Then, in each bundle with two primary items, one item can have payment arbitrarily close to~1, and the other $\frac{1}{2}+s$, with $s$ as above. Hence, to reach $\sim(z-1)$, the bundle needs $\sim(z-1-\frac{3}{2})-s$. As integrating over \emph{all} $s$ we get $q$, the above considerations give the constraint $(6/5)y\ge(z-1-\frac{3}{2})\cdot(3-2x_{1}-3x_{2}-4x_{3}-5x_{4})-q$ as desired. 

In the second case, the number of {primary} items is greater than $3n$. {If $z > 2.5$, then the right-hand side of constraint~4 is a product of a positive scalar and a negative scalar, resulting in a negative scalar. By non-negativity of variables $q$ and $y$, the left-hand side of the constraint is a non-negative scalar. Hence constraint~4 is valid for the range of values of $z$ which will be considered in the proof, which only includes values larger than $2.5$.}

}
}
\end{enumerate}

{The fourth constraint is not a linear inequality. Nevertheless we may make use of the system of four constraints as if it is a system of linear inequalities. We do so by substituting candidate values for $z$ (these values are larger than $2.5$, so that the fourth constraint remains valid), simplifying the second and fourth constraints after making this substitution, and checking whether the resulting system of inequalities is feasible. If it is not feasible, this certifies that the substituted value for $z$ was too high, and hence we obtain an upper bound on $z$. Substituting $z = 2.7$, the constraints become:

\begin{enumerate}
\item $x_{1}+x_{2}+x_{3}+x_{4}+y \leq 1 - \frac{1}{n}$
\item $- 2x_{1}-\frac{3}{2}x_{2}-\frac{4}{3}x_{3}-\frac{5}{4}x_{4}-\frac{6}{5}y+q \leq -1.7$
\item $2x_{1} \leq 1$
\item $-\frac{2}{5}x_{1}-\frac{3}{5}x_{2}-\frac{4}{5}x_{3}-x_{4}-\frac{6}{5}y-q \leq -\frac{3}{5}$. 
\end{enumerate}

Summing up the four constraints multiplied by coefficients $(1.8,1,0.2,0.5)$ 
respectively, we obtain:

$$(\frac{2}{5} - \frac{1}{3})x_3 + \frac{x_4}{20} + \frac{q}{2}  \le -\frac{1.8}{n}$$

As $x_3$, $x_4$ and $q$ are non-negative, this is a contradiction. Hence $z < 2.7$. In fact, the term $-\frac{1.8}{n}$ on the right hand side implies that $z \le 2.7 - \Omega(\frac{1}{n})$, which in turn implies that $\rho \ge \frac{10}{27} + \Omega(\frac{1}{n})$.}
\end{proof} 

\subsection{Polynomial time algorithms}
\label{sec:polynomial}


Theorems~\ref{thm:1/3_APS_guarantee} and~\ref{thm:equal-10/27} imply (among other things) the existence of allocations that give each agent with a submodular valuation a certain fraction of her APS (or MMS). However, they do not provide polynomial time algorithms to find such allocations, because they assume that the value of the APS (or MMS) is known (or can be computed by the agent), whereas computing this value is NP-hard. Nevertheless, by using a technique presented in \cite{DBLP:conf/sigecom/GhodsiHSSY18}, we obtain a polynomial time implementation, proving \Cref{cor:polyTimeSubmodular}. The basic idea is as follows. One runs the bidding game with all agents using our proposed proportional strategy, but each agent starts with an estimate for her MMS (or APS) that is higher than the true value. If all agents get the desired fraction of their estimated MMS, we are done. If not, then for those agents that get a fraction that is too small, we lower their estimate for their MMS by a factor of $(1 - \epsilon)$, and repeat the whole process. No agent will ever need to lower her estimate to below a $(1 - \epsilon)$ fraction of her true MMS. We now provide more details.


Consider an allocation instance $I$ with $n$ agents, $m$ items, integer valued 
valuation functions, with values of bundles in the range $[0,K]$ (that is, $K=\max_{i\in\mathcal{N}}v_i(\items)$). (Alternatively, for a valuation function $v_i$ that is not integer valued, $K$ denotes an upper bound on  $\frac{v_i(\items)}{v_i(S)}$, over all sets $S \subset \items$ for which $v_i(S) > 0$. Namely, $K$ is an upper bound on the ratio between the values of the most valuable bundle and least valuable bundle of positive value.) We assume that all allocation algorithms have value query access to the valuation functions of the agents. We say that an allocation algorithm runs in polynomial time if both its running time and the number of value queries that it makes are polynomial in $n$, $m$ and $\log K$. In the following presentation, we consider the APS as our share notion, and allow for agents of arbitrary entitlement. We remark that the same principles apply (with straightforward modifications) to settings with equal entitlement, and the MMS as a fairness notion. 

\begin{remark}
    Definition~\ref{def:conditional algorithm} and Theorem~\ref{thm:polytime} (that will be presented shortly) involve an approximation ratio $\rho$. All results extend without any change in the proofs to settings in which $\rho$ is not a fixed constant, but instead a function of the entitlement (such approximation ratios appear in Theorem~\ref{thm:1/3_APS_guarantee}, for example). That is, for agent $i$ with entitlement $b_i$, the approximation ratio is $\rho(b_i)$. 
\end{remark}

\begin{definition}
\label{def:conditional algorithm}
For $\rho > 0$, we say that an allocation algorithm is a $\rho$-APS algorithm for a class $C$ of valuations, if given any allocation instance with valuations from the class $C$, the algorithm outputs an allocation $(A_1, \ldots, A_n)$ in which every agent $i$ gets a bundle $A_i$ of value $v_i(A_i) \ge \rho \cdot APS_i$. 
We say that an allocation algorithm is a {\em conditional} $\rho$-APS algorithm for a class $C$ of valuations, if given any allocation instance with valuations from the class $C$, and given any vector $(t_1, \ldots, t_n)$, the algorithm outputs an allocation $(A_1, \ldots, A_n)$ that satisfies the following property: for every agent $i$, if it happens that $t_i \le APS_i$, then $v_i(A_i) \ge \rho \cdot t_i$. 
\end{definition}


The proof of the following theorem is similar to a proof of a related theorem proved in~\cite{DBLP:conf/sigecom/GhodsiHSSY18}. We present its proof for completeness.

\begin{theorem}
\label{thm:polytime}
    Fix arbitrary $\rho > 0$ and arbitrary $\varepsilon > 0$. Then every polynomial time conditional $\rho$-APS algorithm for a class $C$ of valuations can be transformed into a polynomial time (unconditional) $(1 - \varepsilon)\rho$-APS  algorithm for the class $C$ of valuations. Here, the dependence on $\varepsilon$ of the running time of the unconditional algorithm is a multiplicative factor of $O(\frac{1}{\varepsilon})$. 
\end{theorem}

\begin{proof}
{In Algorithm~\ref{alg:conditional to unconditional algorithm} we give an unconditional $(1 - \varepsilon)\rho$-APS algorithm, using a conditional $\rho$-APS algorithm as a blackbox.}

\RestyleAlgo{ruled}
\SetKwComment{Comment}{/* }{ */}
\begin{algorithm}[hbt!]
\caption{An \emph{unconditional} $(\rho - \varepsilon)$-APS algorithm using a \emph{conditional} $rho$-APS algorithm as a blackbox }\label{alg:conditional to unconditional algorithm}
\KwData{$\mathcal{M},\mathcal{N},\langle v_1,\dots,v_n\rangle,  \varepsilon{,K}$}
For every $i\in\mathcal{N},\quad t_i\gets v_i(\items)$\;

\While{true}{
    Run \emph{conditional}-$\rho$-APS algorithm with guesses $\langle t_1,\dots,t_n\rangle$, resulting $\mathcal{A}=\langle A_1,\dots,A_n\rangle$\; 
  \eIf{$\exists i,$ such that $v_i(A_i)<\rho_i\cdot t_i$ {{\bf and} $t_i\geq v_i(\items)\cdot\frac{1}{K}$}}{
    $i = i$ which satisfies the condition\;
    $t_i\gets (1 -\varepsilon)t_i$\;
  }{ Return $\mathcal{A}$ and exit\;
  }
}
\end{algorithm}

{
\begin{remark}
    In Algorithm~\ref{alg:conditional to unconditional algorithm}, we require having $K$ (the maximum ratio between largest and smallest value bundles) as an input. However, in the case of agents with submodular valuations, $K$ is not needed as input. Instead,  $K$ can be computed efficiently as it equals $\max_{i\in\mathcal{N}}\{\max_{\{e\in\mathcal{M}\mid v_i(e)>0\}}\{\frac{v_i(\items)}{v_i(e)}\}\}$. 
\end{remark}
}

\begin{claim}
\label{lower bound of guesses of APS}
During the whole run of Algorithm~\ref{alg:conditional to unconditional algorithm}, for every agent $i$, $t_i\geq (1-\varepsilon)APS_i$.

\end{claim}

\begin{proof}
    Fix an agent $i$. At the beginning of the algorithm, $t_i=v_i(\items)\geq APS_i$. during the algorithm we only reduce the value of $t_i$ each time by a factor of $(1-\varepsilon)$. Consider the first time when $t_i<APS_i$. Then, $t_i\geq (1-\varepsilon)APS_i$ (since in its previous value, the variable $t_i$ was greater than $APS_i$).
    Since $t_i<APS_i$, by \Cref{def:conditional algorithm}, every time we run command 3, the bundle of agent $i$ in the resulted allocation $\mathcal{A}$ is of value $\geq\rho t_i$, so we will not reduce the value of $t_i$.
\end{proof}

We prove the correctness of Algorithm~\ref{alg:conditional to unconditional algorithm}. If the algorithm terminates, then it returns an allocation $\mathcal{A}=\langle A_1,\dots,A_n\rangle$  with the property that for every agent $i$ { with $t_i\geq v_i(\items)\cdot\frac{1}{K}$}, $v_i(A_i)\geq \rho_i \cdot t_i$. By \Cref{lower bound of guesses of APS}, we have that each variable $t_i$ holds during the algorithm a value greater than $(1-\varepsilon) APS_i$. Thus, given that the algorithm terminates, {each such} agent $i$ gets a bundle $A_i$ of value at least $(1-\varepsilon) \rho APS_i$. 
It remains to consider those agents $i$ with $t_i<v_i(\items)\cdot\frac{1}{K}$, and for which furthermore, $APS_i > 0$. The assumption that $APS_i>0$ implies that $APS_i \ge v_i(S)$, where $S$ is the bundle of minimum value. Also, $t_i<v_i(\items)\cdot\frac{1}{K} \le v_i(S)$. 
Thus, by \Cref{lower bound of guesses of APS}, $v_i(A_i)\geq \rho t_i\geq v_i(S) \ge t_i$ (where the middle inequality holds because every bundle of positive value has value at least $v_i(S)$). As $APS_i < \frac{t_i}{1-\varepsilon}$ (otherwise, the condition of step~4 of the algorithm would not allow the value $t_i$ to be reached), we have that $v_i(A_i)\geq (1 - \varepsilon)APS_i$.

We turn to analyze algorithm's time complexity. 
Let $T(n,m,\log K)$ denote the running time of the \emph{conditional}-$\rho$-APS algorithm (called in command 3 of Algorithm ~\ref{alg:conditional to unconditional algorithm}). The number of times that command~3 is run is at most $n\cdot\log_{(1-\varepsilon)}K=n\cdot\frac{1}{\varepsilon}\cdot\log K$ (because in each iteration, at least one $t_i$ drops by a factor of $1-\epsilon$, and the total drop in value of $t_i$ is at most a factor of $K$).
Thus the overall runtime of Algorithm ~\ref{alg:conditional to unconditional algorithm} is $O\left( n\cdot\log(K)\cdot \frac{1}{\varepsilon}\cdot T(n,m,\log K)\right)$, yielding a polynomial time algorithm (under the assumption that $T(n,m,\log K)$ is polynomial).

\end{proof}


The following theorem, Theorem~\ref{thm:conditionalSubmodular}, is a relatively straightforward consequence of Theorems~\ref{thm:1/3_APS_guarantee} and~\ref{thm:equal-10/27}.

\begin{theorem}
\label{thm:conditionalSubmodular}
    There is a polynomial time conditional $\rho$-APS algorithm for submodular valuations, with $\rho = \frac{1}{3 - 2b_i}$. For the equal entitlement case, there is a polynomial time conditional $\rho$-MMS algorithm for submodular valuations, with $\rho = \frac{10}{27} + {\Omega(\frac{1}{n})}$.
\end{theorem}

\begin{proof}
As described in the \emph{proportional bidding strategy}, both in the original and \emph{altruistic} version of the bidding game, an agent $i$ executes the \emph{proportional bidding strategy} is required to know/compute their APS (MMS). Based on her $APS_i$ ($MMS_i$), the agent knows how to bid. Consider a modification of the bidding strategy, in which agent $i$ receives as an auxiliary input a value $t_i$ (instead of computing her true APS (MMS) value), and truncates her valuation function at $t_i$, namely $v_i\gets v_i^{t_i}$ (\Cref{def:$v^t_p$}). Then, the agent infers her bidding using the value of $t_i$ instead of $APS_i$ ($MMS_i$). 

The conditional $\rho$-APS algorithm (or $\rho$-MMS algorithm) is simply to simulate the bidding game (or altruistic bidding game) with the $t_i$ values as auxiliary inputs to the agents, and having each agent follow her respective modified proportional bidding strategy.  Now, giving value queries, the bidding game is simulated in polynomial time, where we break ties arbitrarily. It remains to show the correctness of the algorithm. For this, notice that given a value $t_i\leq APS_i$ ($MMS_i$), then by \Cref{claim:truncation} the $APS$ ($MMS$) value of agent $i$ is reduced to $t_i$, and the truncation preserves submodularity. Now the conditions of \Cref{thm:1/3_APS_guarantee} (\Cref{thm:equal-10/27}) are met, and agent $i$ gets a bundle of value at least $\rho$-APS (-MMS), as desired.
\end{proof}

Combining theorems~\ref{thm:polytime} and~\ref{thm:conditionalSubmodular} we prove \Cref{cor:polyTimeSubmodular}, which is restated here for convenience.

\PolySubmodular*

\begin{proof}
\Cref{thm:conditionalSubmodular} states that there is a conditional $\rho$-APS algorithm for submodular valuations with $\rho=\frac{1}{3-2b_i}$. We can assume $b_i\geq\frac{1}{m}$ (as otherwise the APS of agent $i$ is~0,
implying that $\rho = \frac{1}{3} + \Omega(\frac{1}{m})$. Thus, setting $\varepsilon$ to be $O(\frac{1}{m})$ in Algorithm~\ref{alg:conditional to unconditional algorithm} yields the existence of an unconditional $\frac{1}{3}$-APS polynomial time algorithm, as desired. In a similar way for the $MMS$, by setting $\varepsilon$ to be $O(\frac{1}{n})$ we obtain an unconditional $\frac{10}{27}$-MMS polynomial time algorithm.
\end{proof}



Finally, we restate and prove Corollary~\ref{cor:submodular, additive, and unit demand algorithm}.
\PolyEnsamble*

\begin{proof}
For the sake of this proof, we  consider a modified version of the original bidding game. When an agent wins a round (i.e., she is the highest bidder), instead of picking one item, she can pick $k\geq1$ of the remaining items and pay $k$ times her bid.
We consider this version of bidding game because this is the version for which it was previously shown (in~\cite{BEF21}) that additive agents have a bidding strategy that guarantees themselves $\frac{3}{5}$ of their APS. That strategy is implementable in polynomial time. 

We show that in this version of the bidding game, any submodular, additive, or unit-demand agent has a strategy that guarantees herself the relevant guarantee of her APS.
\begin{itemize}
    \item Submodular agents have a strategy that guarantees $\frac{1}{3}+\Omega(\frac{1}{n})$ fraction of their APS. The key property that enables the proportional bidding strategy presented in \Cref{thm:1/3_APS_guarantee} to maintain its guarantee also in the current version of he bidding game is the fact the bidding sequence of a submodular agent executing the proportional bidding strategy is non-increasing. (\Cref{obsrv: bidding sequence decreasing}).
     Therefore, if another agent, $o$, wins a round and decides to pick $k>1$ items, for the submodular agent perspective, this is equivalent to $k$ rounds in which $o$ raises the same bid, and all other agents raise a bid of zero. Since the bids of the submodular agent are non-increasing, the submodular agent will not raise a bid strictly greater than $o$'s bid. Thus, with adversarial tie-breaking, we can guarantee that in both cases of the bidding game, the bundle of the submodular agent will remain the same, and the $\frac{1}{3}+\Omega(\frac{1}{n})$ guarantee from \Cref{thm:1/3_APS_guarantee} holds.
    
    \item Additive agents have a strategy that guarantees a $\frac{3}{5}$ fraction of their APS. The proof is presented in \cite{BEF21}. In addition, \cite{BEF21} presented a polynomial time implementation of this strategy, that does not require knowing the APS value of an agent.
    
    \item Unit-demand agents have a strategy that guarantees a $1$-APS. Consider a unit-demand agent $p$ with entitlement $b_p$. It is easy to verify that the APS of $p$ is the $\lceil \frac{1}{b_p} \rceil$th most valuable item, and $0$ if there are fewer than $\lceil \frac{1}{b_p} \rceil$ items.
    We claim that the simple strategy of $p$ bidding her entire budget in each round and, upon winning, taking the most valuable remaining item guarantees $p$ her APS.
    Since $p$ bids her entire budget until she wins an item, in every round she does not win, at least $b_p$ of the total budget of agents is spent. Hence, $p$ must win one of the first $\lfloor\frac{1}{b_p}\rfloor$ rounds. By that, she guarantees herself one of the $\lfloor\frac{1}{b_p}\rfloor$ most valuable items, which is at least her APS. Note that this strategy does not require agent $p$ to know her APS value, and given access to value queries, her strategy is polynomial-time.
\end{itemize}

It remains to show a transformation from the existence of approximate fair allocation (induced by the above strategies) to a polynomial time algorithm. The proof of this is similar to the proof of \Cref{cor:polyTimeSubmodular}, and is omitted. 
\end{proof}


\subsection{Negative examples}

{
\begin{proposition}
\label{hard instance altruistic version}
For every constant $\rho > \lim_{k\to\infty}\rho_k \simeq 0.3716$ (where for each $k\in\mathbb{N}$ we will define $\rho_k$ in the proof), there is an allocation instance with equal entitlements and an adversarial run of the \emph{altruistic} version of bidding game, in which an agent $p$ that has a submodular valuation function and uses the proportional bidding strategy gets a bundle of value smaller than $\rho MMS_p$.
\end{proposition}

\begin{proof}
We present a series of instances in which agent
$p$ with a submodular valuation function executes the proportional bidding strategy.

The instances are parameterized by $k\in\mathbb{N}$. The $k$th instance will be as follows:
Define
\begin{align*}
q_{1} & =2\\
q_{k} & =1+\prod_{i=1}^{k-1}q_{i}
\end{align*}

(This sequence is known as the Sylvester sequence)

The number of agents will be: $n_{k}=q_{k+1}-1$ (for example, for
$k=2$, $n_{k}=2\cdot3\cdot7=43$).

The set of item is $\mathcal{M}=\{e_{i,j}\}$ for $1\leq i\leq k+1$,
$1\leq j\leq n$ ($n\cdot(k+1)$ items).

If we think of $e_{i,j}$ as arranged in a matrix, then all the items
in a row are copies of the same item and are substitutes. The value
of items from different rows is additive.

For any $1\le i\leq k$ and for any $j$, $v_{p}(e_{i,j})=\frac{1}{q_i-1}$.
For $i=k+1$ and any $j$, $v_{p}(e_{k+1,j})=1$. For example, if $k=3$,
there are $43$ agents and columns, and in each column $j$, $v_p(e_{1,j})=1$, $v_p(e_{2,j})=\frac{1}{2}$, $v_p(e_{3,j})=\frac{1}{6}$, $v_p(e_{4,j})=1$.

\begin{itemize}

\item $v_{p}$ is submodular. The marginal value of each item is weakly decreasing (the marginal value of item $e_{i,j}$ to a set $S$ is either $v_{p}(e_{i,j})$ or $0$, depending on whether the set $S$ already contains an item from the $i$'th row).

\item The columns $C_j$ of the matrix $\{e_{i,j}\}$ form an MMS partition. The value of every bundle is at most $v_{p}(\mathcal{M})$, and in this partition, the value of each bundle (column) is exactly $v_{p}(\mathcal{M})$.

\item $APS_p = MMS_p =  v_p(\mathcal{M})=v_{p}(C_j)=v_p(e_{k+1,j})+\sum_{i=1}^{k}v_{p}(e_{i,j})=1+\sum_{i=1}^{k}\frac{1}{q_i-1}$

\item $q_{i}$ divides $n$, for every $i \le k$.

\end{itemize}

For convenience, we assume w.l.o.g that the budget of each agent equal her $MMS$.

For every $k$, we first show a run of the bidding game with adversarial bidding of the other agents, in which agent $p$ executes the proportional bidding strategy with $\rho_{k}=\frac{1}{MMS_p}=\frac{1}{1+\sum_{i=1}^{k}\frac{1}{q_i-1}}$, and she receives a value of precisely $1$ (she gets the bundle that consists only of items from row $k+1$), which is a $\frac{1}{MMS_{p}}$ of her $MMS$.
The series of $\rho_{k}$ is monotonically decreasing and bounded by 0, so $\lim_{k\to\infty}\rho_k$ exists. (Sylvester's sequence grows at a doubly exponential rate. Hence, the sequence of $\rho_{k}$ converges very fast.)


Then, by $proportional(\rho_{k})$, in each round, agent $p$ bids the highest marginal value of the remaining items. Moreover, each agent who spends more than $1$ value from her budget becomes inactive.
We now present the adversarial run.

In round $1$, $p$ bids $1$, and is allowed to win. She selects an item of value $1$ from row $k+1$. 

In each of the next $n$ rounds, at least one of the first $\frac{n}{2}$ other agents bids $1$, and upon winning (note that $p$ bids $1$ in each of these rounds, and the algorithm is assumed to brake the ties adversarially), takes an item from the first row (i.e., $e_{1,j}$). 
All items of the first row are taken by $\frac{n}{2}$ of the other agents. These $\frac{n}{2}$ agents surpass a payment of $1$ and become inactive. 


In each of the next $n$ rounds, at least one of the next $\frac{n}{3}$ other agents bids $\frac{1}{2}$, and upon winning (note that $p$ bids $\frac{1}{2}$ in each of these rounds), takes an item from the second row (i.e., $e_{1,j})$. Each such agent surpasses a payment of 1 exactly when winning her 3rd item, and becomes inactive.

The run proceeds in the same way, where for every $i$, $\frac{n}{q_{i}}$ of the other agents bid $\frac{1}{q_{i}-1}$, win all the items in the $i$'th row, and become inactive. Note that  each such agent surpasses a payment of $1$ exactly when winning her $q_i$th item, and becomes inactive. Note that we use the property of $q_{i}\mid n$ for every $i\leq k$.

Thus, the number of other agents that take all items from rows $1$ to $k$ is:
\[
\sum_{i=1}^{k}\frac{n}{q_{i}}=n\cdot\sum_{i=1}^{k}\frac{1}{q_{i}}=n\cdot(1-\frac{1}{q_{k+1}-1})=n\cdot(1-\frac{1}{n})=n-1
\]

Thus, there are sufficiently many other agents to take all items from rows $1$ to $k$, and agent $p$ gets items only from row $k+1$. As they are substitutes, the total value received by $p$ is $1$.

Consider for $\rho'>\rho_{k}$ the altruistic version of the bidding game in which an agent becomes inactive after spending a $\rho'$ fraction of her budget, and suppose that $p$ executes the $proportional(\rho')$ bidding strategy. 
On the instance $I_k$ described above, the same run of the bidding game holds, and $p$ does not get a bundle of value $\rho' APS_p$, but rather only $\rho_k APS_p$. Hence, $I_{k}$ serves as an example showing that our proof of Theorem~\ref{thm:1/3_APS_guarantee} does not extend to values of $\rho$ larger than $\rho_{k}$. The same holds for every $\rho>\lim_{k\to\infty}\rho_k$ (by enlarging $k$, we can make $\rho_k$ as close as we wish to $\lim_{k\to\infty}\rho_k$).
\end{proof}
}

\begin{remark}
The negative example in Proposition~\ref{hard instance altruistic version} can be modified by replacing (for every $j$) the single item $e_{k+1,j}$ of value~1 by $q_k - 1$ items, each of value $\frac{1}{q_k - 1}$ (the same value as that of item $e_{k,j}$). In this modified version, the adversarial run is changed so that other agents win all items $e_{ij}$ for $i \le k$ (their budgets exactly suffice for this), and agent $p$ can take the remaining items from one of the bundles of the MMS partition (the remaining items in different MMS bundles are substitutes to each other and do not provide additional marginal value), getting a value of~1. This modified example is useful in illustrating that for certain variations of the bidding game (considered by the authors but omitted here), bidding strategies similar to the ones considered in the proof of Theorem~\ref{thm:equal-10/27}  do not lead to approximation ratios that are significantly better than those proved in Theorem~\ref{thm:equal-10/27}).  
\end{remark}



The following proposition shows that in \Cref{thm:1/3_APS_guarantee}, the value of $\rho$ cannot be improved to a constant (independent of $b_p$) larger than $\frac{1}{3}$. Its proof is similar to the proof of Proposition~\ref{hard instance altruistic version}, with some relatively straightforward modifications. For completeness, its full proof is presented in the appendix (Section~\ref{sec:example}).

\begin{restatable}{reprop}{exampleThird}
\label{no rho larger than 1/3 for original bidding game}
For every constant $\rho > \frac{1}{3}$, there is an allocation instance with equal entitlements and an adversarial run of the bidding game, in which an agent $p$ that has a submodular valuation function and uses the $proportional(\rho)$ bidding strategy gets a bundle of value smaller than $\rho MMS_p$.
\end{restatable}





We now restate and prove \Cref{prop:XOS_hardness}, showing that our proof techniques do not extend to XOS valuations.
\XosHardness*

\begin{proof}
For parameters $n,k$, define the instance $I(n,k)$ as follows. There
are $n$ agents with equal entitlements. The set $\mathcal{M}$ of items  consists of $nk$ items $e_{ij}$ for $1\leq i\leq k$ and $1\leq j\leq n$. We think of $\mathcal{M}$ as arranged in an $k\times n$ matrix, with $e_{ij}$ in the $ij$ entry. For every column $j$, let $c_{j}$ be the additive valuation function defined by giving value $1$ to items in column $j$, and $0$ to all other items. Let $v$ be the pointwise maximum of the functions $c_{j}$, that is, $v(S)=\max_{j}c_{j}(S)$ for every $S\subseteq\mathcal{M}$.
Then the valuation function $v$ is an XOS function by its definition.
We focus on a specific agent $p$ whose valuation function is $v_{p}=v$.
(The other agents may have arbitrary valuations.)

\begin{claim}
\label{claim:XOS_hardness}
If $n\geq4k^{2}$, no bidding strategy can guarantee $p$ more than a $1/k$-fraction
of $MMS_{p}$.
\end{claim}

\begin{proof}
For convenience, assume all agents are given a budget of $k$. We give
the other agents adversarial bidding strategies, as follows. There
are two types of agents.
\begin{itemize}
\item Type 1 agents consist of $n/2$ of the agents that always bid $1/2$,
and take an arbitrary available item upon winning.
\item Type 2 agents consist of the rest $n/2-1$ agents, which operate
as follows. Once agent $p$ wins an item $e_{ij}$, an agent of this
type bids all of her budget in the next $k-1$ rounds, and upon winning,
chooses an available item from column $j$ (and becomes inactive).
\end{itemize}
An agent of type 1 becomes inactive after winning exactly $2k$ items.
As the number of items is $nk$, it follows that there exists an active
agent of type 1 in every round. Thus, agent $p$ must pay $1/2$ for
every won item, so she can win at most $2k$ items overall. Once $p$
wins her first item from some column $j$, if there exist at least
$k-1$ active agents of type 2, all other items in column $j$ will
be taken by them in the next $k-1$ rounds. So, if we start with at
least $(2k)\cdot(k-1)$ agents of type 2, $p$ will not win more than
one item from every column. As $(2k)\cdot(k-1)\leq2k^{2}-1\leq n/2-1$,
this indeed holds, so agent $p$ cannot win a bundle of value more
than $1$. Observe that $MMS_{p}=k$, so the claim follows.
\end{proof}

Proposition~\ref{prop:XOS_hardness} is an immediate consequence of Claim~\ref{claim:XOS_hardness}.
\end{proof}

\bibliographystyle{alpha}
\bibliography{Refrences_DB}

\begin{appendix}

\section{APX-hardness for computing MMS and APS for submodular valuations}
\label{sec:APX}

The current section is presented in a somewhat sketchy way, and without detailed proofs, as it is not the main focus of the current paper.

In general, computing the MMS and the APS are both NP-hard tasks. For example, for the case of two agents with equal entitlements and additive valuations, weak NP-hardness is a straightforward consequence of the NP-hardness of the PARTITION problem (given a set of integers, is there a subset whose sum of values is exactly half of the total sum?). Computing the APS is in some sense an easier task than computing the MMS. In particular, for additive valuations there are pseudo-polynomial time algorithms for computing the APS~\cite{BEF21}, whereas computing the MMS is strongly NP-hard. Also, as the value of the APS is a solution to a linear program with exponentially many constraints (Definition~\ref{def:APS}), the APS can be computed in polynomial time (using the ellipsoid algorithm) if there is a separation oracle for the linear program. This separation oracle corresponds to a computational problem that has a natural economic interpretation: given prices to the items and a budget for the agent, which is the highest value bundle that the agent can afford? For a given valuation function, if one can answer such queries in polynomial time, then the APS can be computed in polynomial time. 

In the case of equal entitlement, the APS is at least as large as the MMS, and sometimes strictly larger. For submodular valuations, a ratio of $\frac{5}{6}$ between the MMS and the APS is demonstrated in~\cite{BEF21}, for an allocation instance with six items and two agents with equal entitlements.

For submodular valuations, both the MMS and the APS are APX-hard to compute. We are not aware of a reference in which such a statement is proved explicitly. However, $(1 - \frac{1}{e})$ approximation hardness can be proved using known techniques. (For simplicity, we omit here low order additive terms when stating approximation ratios.) Let us briefly explain how. A certain reduction template (starting from the APX-hard problem Max 3SAT) described in~\cite{Feige98} established hardness of approximation results for Min Set Cover  (within a ratio of $\ln n$) and Max $k$-Coverage (within a ratio of $1 - \frac{1}{e}$). The same reduction template, but starting from a different APX-hard problem (Max 3-Coloring), was used in~\cite{Feige_domatic02} to prove hardness of approximation of the Domatic Number (within a ratio of $\ln n$), and  in~\cite{DBLP:journals/algorithmica/KhotLMM08} to prove that the maximum welfare problem with submodular valuations is hard to approximate within a ratio of $1 - \frac{1}{e}$. The reduction used to prove this last result implies that for submodular valuations it is NP-hard to approximate the MMS within a ratio better than $1 - \frac{1}{e}$. The reason for this is that in the maximum welfare instance constructed by the reduction, all agents have the same submodular valuation function (call it $v$). Moreover, on {\em yes} instances, the maximum welfare allocation gives all agents the same value (call it $t$, with the total welfare being $n\cdot t$). Hence on {\em yes} instances, the MMS of $v$ is $t$ (as the maximum welfare allocation serves as an MMS partition). On {\em no} instances, the fact that the welfare is at most $(1-e)n\cdot t$ implies that in every partition to $n$ bundles, at least one of the bundles has value at most $(1-e)t$, showing that the MMS is at most $(1-\frac{1}{e})t$. As it is NP-hard to distinguish between {\em yes} and {\em no} instances, we get a hardness of approximation for the MMS.

To derive $(1 - \frac{1}{e})$ approximation hardness for the APS, one further observes the following property of {\em no} instances that result from using the reduction template: every set that contains a $\frac{1}{n}$ fraction of the items has value at most $(1 - \frac{1}{e})t$. This property is inherited from the fact that the same reduction template is used in the proof in~\cite{Feige98} that Max $k$-Coverage is hard to approximate within a ratio better than $1 - \frac{1}{e}$. When the entitlement is $\frac{1}{n}$, the APS fractional partition (of Definition~\ref{def:dual}) must contain at least one bundle with at most  a $\frac{1}{n}$ fraction of the items, and hence the APS for {\em no} instances is at most $(1 - \frac{1}{e})t$.


\section{A simple technical claim}


{We recall notation used when describing the $proportional(\rho)$ bidding strategy for the bidding game in Section~\ref{sec:APS}. Let $I_{s}$ denote the residual instance after round $s$, which is the first point in time after which no items satisfy $v_{p}(e)>2\rho APS_{p}$. The total budget $b^s$ remaining for all agents is denoted by $\gamma$. We consider a new instance
$\hat{I_{s}}$ to be the following:
\begin{itemize}
\item The set of agents is those who are active in $I_{s}$.
\item The entitlement of an agent in $\hat{I_{s}}$ is her budget in
$I_{s}$, scaled by $\frac{1}{\gamma}$. Consequently, the entitlements
are non-negative and sum up to 1.
\item The items of $\hat{I_{s}}$ are those of $I_{s}$ (denoted by $\items^s$), and $v_p$ remains unchanged (over subsets of $\items^s$).
\end{itemize}}

{
\begin{claim}
\label{claim:APS not decrease in I_s}
Given that agent $p$ did not win an item yet (in the first $s$ rounds),
then $APS(\mathcal{M}^{s},v_{p},\frac{1}{\gamma}b_{p})\geq APS(\mathcal{M},v_{p},b_{p})$.
In other words, the $APS$ of agent $p$ in $\hat{I_{s}}$ is at least
as her APS in $I_{0}$, the original instance. (Referring \Cref{def:$v^t_p$}, in the special case of $v_p=v_p^t$ with $t=APS_p$, then the claim holds with equality)
\end{claim}
\begin{proof}
Agent $p$ did not win an item in the first $s$ rounds, while there
is an item with a value greater than $2\rho$ of her APS. Therefore
by the proportional bidding strategy, she bids her entire budget,
$b_{p}$, in each of these rounds. Thus, after these $s$ rounds,
$\gamma=b^{s}\geq1-s\cdot b_{p}$. Let $P_{0}=\{\lambda_{S}S\}$ be
a fractional partition associated with the APS of agent $p$. then
we define a new fractional partition $P_{s}=\{\lambda'_{S}S\}$ to
be the following
\[
\lambda'_{S}=\begin{cases}
\frac{1}{\gamma}\lambda_{S} & \text{ if }S\subseteq\mathcal{M}^{s}\\
0 & \text{otherwise}
\end{cases}
\]

We claim that $P_{s}$ witness that indeed $APS(\mathcal{M}^{s},v_{p},\frac{1}{\gamma}b_{p})\geq APS(\mathcal{M},v_{p},b_{p})$.
\begin{itemize}
\item 
\begin{align*}
\sum_{S\subseteq\mathcal{M}^{s}}\lambda'_{S} & =\frac{1}{\gamma}\sum_{S\subseteq\mathcal{M}^{s}}\lambda{}_{S}=\frac{1}{\gamma}(\sum_{S\subseteq\mathcal{M}}\lambda_{S}-\sum_{e\in\mathcal{M\setminus\mathcal{M}}^{s}}\sum_{S\mid e\in S}\lambda_{S})\\
 & \underset{*}{\geq}\frac{1}{\gamma}(1-\sum_{e\in\mathcal{M\setminus\mathcal{M}}^{s}}b_{p})\\
 & =\frac{1}{\gamma}(1-s\cdot b_{p})\\
 & \geq1
\end{align*}
By definition of $P_o$, for each $e\in\items$, $\sum_{S\mid e\in S}\lambda_S\leq b_p$, which justify inequality *.
\item Each $e\in\mathcal{M}^{s}$ satisfies:
\[
\sum_{S\mid e\in S}\lambda'_{S}=\frac{1}{\gamma}\sum_{S\mid e\in S}\lambda_{S}\le\frac{1}{\gamma}b_{p}
\]
 Where $\frac{1}{\gamma}b_{p}$ is the entitlement of $p$ in $\hat{I_{s}}$.
\item Each $S\subseteq\mathcal{M}^{s}$ with strictly positive weight in $P_s$ is of value $v_{p}(S)\geq APS_{p}$ (since a bundle with positive weight
also has positive weight in $P_0$). In the special case of $v_p=v_p^t$ with $t=APS_p$, (no bundle equals more than $APS_p$) then clearly also $v_p(S)\leq APS_p$, and therefore $v_p(S)=APS_p$
\end{itemize}
\end{proof}
}

\section{A negative example}
\label{sec:example}

We restate and prove \Cref{no rho larger than 1/3 for original bidding game}.

\exampleThird*

\begin{proof}
We present a series of instances in which agent $p$ with a submodular valuation function executes the proportional bidding strategy.

The instances are parameterized by $k\in\mathbb{N}$. The $k$th instance will be as follows:
Define
\begin{align*}
q_{1} & =2\\
q_{k} & =1+\prod_{i=1}^{k-1}q_{i}
\end{align*}

(This sequence is known as the Sylvester sequence)

The number of agents will be: $n_{k}=q_{k+1}-1$ (for example, for
$k=2$, $n_{k}=2\cdot3\cdot7=43$)

The set of item is $\mathcal{M}=\{e_{i,j}\}$ for $1\leq i\leq k+1$,
$1\leq j\leq n$ ($n\cdot(k+1)$ items)

If we think of $e_{i,j}$ as arranged in a matrix, then all the items
in a row are copies of the same item and are substitutes. The value
of items from different rows is additive.

For any $1\le i\leq k$ and for any $j$, $v_{p}(e_{i,j})=\frac{2}{q_{i}}$.
For $i=k+1$ and any $j$, $v_{p}(e_{k+1,j})=1$. For example, if $k=3$,
there are $43$ agents and columns, and in each column $j$, $v_{p}(e_{1,j})=1$, $v_{p}(e_{2,j})=\frac{2}{3}$, $v_{p}(e_{3,j})=\frac{2}{7}$, $v_{p}(e_{4,j})=1$.

\begin{itemize}

\item $v_{p}$ is submodular. The marginal value of each item is weakly decreasing (the marginal value of item $e_{i,j}$ to a set $S$ is either $v_{p}(e_{i,j})$ or $0$, depending on whether the set $S$ already contains an item from the $i$'th row).

\item The columns $C_j$ of the matrix $\{e_{i,j}\}$ form an MMS partition. The value of every bundle is at most $v_{p}(\mathcal{M})$, and in this partition, the value of each bundle (column) is exactly $v_{p}(\mathcal{M})$.

\item $APS_p = MMS_{p} =  v_{p}(\mathcal{M})=v_{p}(C_{j})=v_{p}(e_{k+1,j})+\sum_{i=1}^{k}v_{p}(e_{i,j})=1+\sum_{i=1}^{k}\frac{2}{q_{i}}=1+2\sum_{i=1}^{k}\frac{1}{q_{i}}=1+2\cdot(1-\frac{1}{q_{k+1}-1})\underset{*}{=}3-\frac{2}{q_{k+1}-1}$,
where equality {*} is a known property of the partial sums of Sylvester's inverse series (this can be proved by induction, Wikipedia value of Sylvester sequence).

\item $q_{i}$ divides $n$, for every $i \le k$.

\end{itemize}

For convenience, we assume w.l.o.g that the budget of each agent is $2$.

For every $k$, we first show a run of the bidding game with adversarial bidding of the other agents, in which agent $p$ executes the proportional bidding strategy with $\rho_{k}=\frac{1}{APS_{p}}$, and she receives a value of precisely $1$ (she gets the bundle that consists only of items from row $k+1$) which is a $\frac{1}{APS_{p}}$ of her $APS$.
For that instance, $\rho_{k}=\frac{1}{3-\frac{2}{q_{k+1}-1}}$. The series of $\rho_{k}$ is monotonically decreasing to a limit of $\frac{1}{3}$. (Sylvester's sequence grows at a doubly exponential rate. Hence, the sequence of $\rho_{k}$ converges very fast.)


Consider the $I_{k}$ instance parameterized by $k$. For convenience, assume the budget of each agent is $2$ (which is $2\rho_{k}APS_{p})$. Then, by $proportional(\rho_{k})$, in each round, agent $p$ bids the highest marginal value of the remained items.
We now present the adversarial run.

In round $1$, $p$ bids $1$, and is allowed to win. She selects an item of value $1$ from row $k+1$. 

In each of the next $n$ rounds, at least one of the first $\frac{n}{2}$ other agents bids $1$, and upon winning (note that $p$ bids $1$ in each of these rounds, and the algorithm is assumed to brake the ties adversarially), takes an item from the first row (i.e., $e_{1,j}$). 
All items of the first row are taken by $\frac{n}{2}$ of the other agents. These $\frac{n}{2}$ agents exhaust their budget and become inactive. 


In each of the next $n$ rounds, at least one of the next $\frac{n}{3}$ other agents bids $\frac{2}{3}$, and upon winning (note that $p$ bids $\frac{2}{3}$ in each of these rounds), takes an item from the second row (i.e., $e_{1,j})$. Each such agent becomes inactive after taking three items.

The run proceeds in the same way, where for every $i$, $\frac{n}{q_{i}}$ of the other agents bid $\frac{2}{q_{i}}$, win all the items in the $i$'th row, and become inactive. Note that we use the property of $q_{i}\mid n$ for every $i\leq k$.

Thus, the number of other agents that take all items from rows $1$ to $k$ is:
\[
\sum_{i=1}^{k}\frac{n}{q_{i}}=n\cdot\sum_{i=1}^{k}\frac{1}{q_{i}}=n\cdot(1-\frac{1}{q_{k+1}-1})=n\cdot(1-\frac{1}{n})=n-1
\]

Thus, there are sufficiently many other agents to take all items from rows $1$ to $k$, and agent $p$ gets items only from row $k+1$. As they are substitutes, the total value received by $p$ is $1$.

Notice that if $p$ executes $proportional(\rho'$) with $\rho'>\rho_{k}$, then the bids of $p$ in each round are strictly smaller than those described above. Hence same run of the algorithm holds, and $p$ does not get a bundle of value $\rho' APS_p$, but rather only $\rho_k APS_p$. Hence, $I_{k}$ serves as an example showing for every $\rho'>\rho_{k}$ that $proportional(\rho')$ does not guarantee $p$ a value of $\rho' APS_p$.

Since $\rho_{k}$ is close to $\frac{1}{3}$ as we wish, for any $\rho>\frac{1}{3}$, there exist a witness $I_{k}$ on which $proportional(\rho)$ does not guarantee $p$ a $(\rho)$-fraction of her $APS$.
\end{proof}

\end{appendix}

\end{document}